\documentclass[aps,preprint,floatfix,nofootinbib,showpacs]{revtex4-1}
\pdfoutput=1
\usepackage{graphicx,color,rotating}
\usepackage{hyperref}
\usepackage{epsfig,color}

\newcommand{\lsim}{\raisebox{-0.13cm}{~\shortstack{$<$ \\[-0.07cm] $\sim$}}~}
\newcommand{\gsim}{\raisebox{-0.13cm}{~\shortstack{$>$ \\[-0.07cm] $\sim$}}~}

\def\beq{\begin{equation}}
\def\eeq{\end{equation}}

\def\bea{\begin{eqnarray}}
\def\eea{\end{eqnarray}}
\def\bei{\begin{itemize}}
\def\eei{\end{itemize}}
\def\bmat{\begin{matrix}}
\def\emat{\end{matrix}}
\def\ble{\begin{flushleft}}
\def\ele{\end{flushleft}}
\def\={\,=\,}
\def\+{\,+\,}
\def\-{\,-\,}

\begin{document}

\def\thefootnote{\fnsymbol{footnote}}

{\small
\begin{flushright}
CNU-HEP-15-04
\end{flushright} }

\title{Bounds on Higgs-Portal models from the LHC Higgs data}
\author{
Kingman Cheung$^{1,2,3}$, P. Ko$^4$,
Jae Sik Lee$^{3,5}$, and
Po-Yan Tseng$^1$}
\affiliation{
$^1$ Department of Physics, National Tsing Hua University,
101, Section 2 Kuang-Fu Road, Hsinchu, Taiwan \\
$^2$ Division of Quantum Phases and Devices, School of Physics,
Konkuk University, Seoul 143-701, Republic of Korea \\
$^3$ Physics Division, National Center for Theoretical Sciences,
101, Section 2 Kuang-Fu Road, Hsinchu, Taiwan\\
$^4$ School of Physics, KIAS, 85 Hoegiro Dongdaemun-gu, Seoul 02455, Korea\\
$^5$ Department of Physics, Chonnam National University, \\
300 Yongbong-dong, Buk-gu, Gwangju, 500-757, Republic of Korea
}
\date{\today}

\begin{abstract}
In a number of Higgs-portal models, an $SU(2)$ isospin-singlet scalar
boson generically appears at the electroweak scale and can mix with
the Standard Model (SM) Higgs boson  with a mixing angle $\alpha$.
This singlet scalar boson can have renormalizable couplings to a pair of dark 
matter particles, vector-like leptons or quarks, or new gauge bosons,  
thereby modifying the Higgs signal strengths in a nontrivial way.  
In this work, we perform global fits to such models using the most updated 
LHC Higgs-boson data and discuss the corresponding implications on
Higgs-portal-type models.  
In particular we find that the current LHC Higgs-boson data slightly 
favors the SM over the Higgs-portal singlet-scalar 
models, which has to be further examined using the upcoming LHC Higgs-boson
data. 
Finally,
without non-SM particles contributing to the
$H\gamma\gamma$ and $Hgg$ vertices, 
the Higgs-portal models 
are constrained as follows:
$\cos\alpha \gtrsim 0.86$ and $\Delta \Gamma_{\rm tot} \lesssim 1.24$ MeV
at 95 \% confidence level (CL). 
\end{abstract}

\maketitle

\section{Introduction}
%

After the discovery of the 125 GeV boson at the LHC \cite{atlas,cms}, 
many analyses have
been performed on the Higgs-boson data in order to identify the nature
of the observe boson. So far, its properties are very close to those
of the Standard Model (SM) Higgs boson: namely, 
(i) the spin, parity, and charge conjugation quantum numbers are 
equal to $J^{PC} = 0^{++}$, which are in accordance with
the SM Higgs boson, and
(ii) its couplings to the SM particles are close to
those of the SM Higgs boson at the end of the LHC Run I,
which is indeed a remarkable achievement.  
The Higgs couplings to the SM
particles are often parameterized in terms of the $\kappa$'s defined
as follows~\cite{LHCHiggsCrossSectionWorkingGroup:2012nn}: \beq
\kappa_i^2 \= \frac{\Gamma(H \to ii)}{\Gamma( H\to ii )_{SM}}, \qquad
\kappa_H^2 \= \frac{ \Gamma_{\rm tot}(H)+\Delta\Gamma_{\rm tot}}{\Gamma_{\rm SM}}\,,
\label{eq:kappa1} 
\eeq
where $i=W,Z,f,g,\gamma$,
and $\Gamma_{\rm SM}$ denotes the SM total decay width while
$\Gamma_{\rm tot}(H)$ and $\Delta\Gamma_{\rm tot}$ 
denote, respectively,  the total decay width into the
SM particles with modified couplings and 
an arbitrary non-SM contribution 
to the total decay width. 
The current best fits to the $\kappa_i$'s  for $i=W,Z,f$  
from the ATLAS \cite{atlas-kappa}
and the CMS \cite{cms-kappa} collaborations are summarized in Table~\ref{table1}.

\begin{table}[th!]
\caption{\small \label{table1}
The best fit values of $\kappa$'s from the ATLAS \cite{atlas-kappa}
and CMS \cite{cms-kappa} at the end of the LHC Run I. The errors or the
ranges are at 68\% CL. }
\medskip
\begin{ruledtabular}
\begin{tabular}{lcccccc}
 & $\kappa_W$ & $\kappa_Z$ & $\kappa_t$ & $\kappa_b$ & $\kappa_\tau$ & $\kappa_\mu$\\
ATLAS & $0.68^{+0.30}_{-0.14}$ & $ 0.95^{+0.24}_{-0.19}$  &
    \scriptsize{$ [-0.80,-0.50] \cup [0.61,0.80]$} & $[-0.7,0.7]$ &
  \scriptsize{ $[-1.15,-0.67] \cup [0.67,1.14]$}  & - \\
CMS & $ 0.95^{+0.14}_{-0.13} $ & $ 1.05^{+0.16}_{-0.16}$ & $0.81^{+0.19}_{-0.15}$ & 
 $0.74^{+0.33}_{-0.29}$ & $0.84^{+0.19}_{-0.18}$ & $0.49^{+1.38}_{-0.49} $ 
\end{tabular}
\end{ruledtabular}
\end{table}

New physics beyond the Standard Model (BSM) 
will be manifest itself if $\kappa_i \neq 1$ for some $i$
in this approach.  Very often it is assumed that the new physics
effects are decoupled from the SM sector, thereby can be described by
nonrenormalizable higher dimensional operators 
\cite{NRop}.  
This assumption
encompasses a large class of BSMs, but still leaves out another large
class of BSMs with an isospin-singlet scalar boson (of a mass around the electroweak
(EW) scale) that could mix with the SM Higgs boson.  This singlet
scalar boson itself can couple to new particles such as a pair of dark
matter (DM) particles, new vector-like quarks and/or leptons, new charged or
neutral vector bosons, etc., just to name a few (see
Ref.~\cite{Chpoi:2013wga} for more comprehensive discussion). Such a
mixing between the singlet scalar boson and the SM Higgs boson
does not decouple and cannot be captured by the usual higher
dimensional operators, and therefore has to be treated in a separate manner.

In Ref.~\cite{Chpoi:2013wga}, a new parameterization was proposed which
is suitable in the presence of a new singlet scalar boson that mixes with the
SM Higgs boson.  The singlet-mixed-in case deserves closer
investigation, because many BSMs with good physics motivations come with
an extra singlet scalar boson that can mix with the SM Higgs
boson. This includes a large class of hidden-sector dark matter
models such as Higgs-portal fermion or vector DM models, and DM models
with local dark gauge symmetries, as well as nonsupersymmetric
$U(1)_{B-L}$ model, vector-like fermions that could affect
$h\rightarrow gg, \gamma\gamma$, or models with the dilaton coupled to the
trace of energy-momentum tensor.

The Higgs-boson properties could be affected by the presence of new physics 
from different origins.  The approach using $\kappa_i$'s is simple and straightforward 
but in general it is difficult to further analyze the origin of new 
physics that had modified the $\kappa$'s from the SM values.
There are basically two different approaches to consider the new physics
effects: one assumes either (i) the full $SU(3)_C \times SU(2)_L \times
U(1)_Y$ gauge symmetry or (ii) its unbroken subgroup $SU(3)_C \times
U(1)_{\rm em} $ only in the effective Lagrangian for Higgs physics.
Though either approach works as long as one is interested in any possible
deviations of the Higgs couplings from the SM values,  
it would be more proper to impose the full gauge symmetry for
investigations at the EW scale, because the energy and momentum
transfer would be $\sim O (m_Z)$ or higher.  On the other hand, if we only 
impose the unbroken subgroup of the SM gauge group, the observed Higgs boson
could be a mixture of the SM Higgs boson and other neutral scalar
bosons that could mix with the SM Higgs boson after electroweak
symmetry breaking (EWSB).  Therefore,  in order to isolate the effects of the mixing 
between the singlet scalar  boson and the SM Higgs boson, we shall impose the full SM 
gauge  symmetry when we construct the effective Lagrangian for the SM Higgs boson
and the singlet scalar boson.

In this paper, we perform the global fits to new physics scenarios with
an extra singlet scalar boson mixed with the SM Higgs boson using the most
recent Higgs data from LHC@7 and 8TeV. In Sec.~II, we set up the
formalism used in this analysis, and compare it with the approach by
the LHC Higgs Cross Section Working Group.  In Sec.~III, we give brief
description of the models which are covered by our formalism.  In
Sec.~IV, we perform the numerical analysis with global fits to the LHC
Higgs data, and present the best $\chi^2$ fit for each model, and
discuss the corresponding implications. Finally we summarize the 
results in Sec.~V.

\section{Formalism}

In the following, we first describe the SM Higgs couplings to 
SM particles including fermions $f$ and gauge bosons $W,Z,\gamma,g$, 
and define
a set of ratios $b_{W,Z,f,\gamma,g}$, which denote the size of the
couplings relative to the corresponding SM one. Without loss of
generality, we define a similar set of ratios $c_{W,Z,f,\gamma,g}$ for the
singlet scalar boson couplings to the fermion $f$ and gauge bosons
$W,Z,\gamma,g$ relative to the corresponding one of the SM Higgs boson.
After then we describe the mixing between the SM Higgs field and
the singlet field via a mixing angle $\alpha$.

\subsection{SM Higgs Couplings}
%

The couplings of the SM Higgs $h$ to fermions are given by
\begin{equation}
{\cal L}_{h\bar{f}f}\ =\ - \sum_{f=u,d,l}\,\frac{g m_f}{2 M_W}\,b_f\,
h\, \bar{f}\,f\,,
\end{equation}
and its couplings to the the massive vector bosons by
\begin{equation}
{\cal L}_{hVV}  =  g\,M_W \, \left(
b_W W^+_\mu W^{- \mu}\ + \
b_Z \frac{1}{2\, \cos^2\theta_W }\,Z_\mu Z^\mu\right) \, h\,,
\end{equation}
where $\theta_W$ is the weak mixing angle.
In the SM limit, we have $b_f=b_W=b_Z=1$. 

While the SM Higgs coupling to two photons is defined through 
the amplitude for the decay process
$h \rightarrow \gamma\gamma$ and it can be written as
\begin{eqnarray} \label{hipp}
{\cal M}_{\gamma\gamma h}=-\frac{\alpha M_{H}^2}{4\pi\,v}\,
S^\gamma_h\,\left(\epsilon^*_{1\perp}\cdot\epsilon^*_{2\perp}\right)
\end{eqnarray}
where $\epsilon^\mu_{1\perp} = \epsilon^\mu_1 - 2k^\mu_1 (k_2 \cdot
\epsilon_1) / M^2_{H}$, $\epsilon^\mu_{2\perp} = \epsilon^\mu_2 -
2k^\mu_2 (k_1 \cdot \epsilon_2) / M^2_{H}$ 
with 
$\epsilon_{1,2}$ being the wave vectors of the two photons and
$k_{1,2}$ being the momenta of the corresponding photons 
with $(k_1+k_2)^2=M_H^2$.
Including some additional loop contributions from non-SM particles
and retaining only the dominant loop
contributions from the third--generation fermions and $W^\pm$,
the scalar form factor is given by
\begin{eqnarray}
S^\gamma_h=2\sum_{f=b,t,\tau} N_C\,
Q_f^2\, b_f\,F_{sf}(\tau_{f}) - b_W\,F_1(\tau_{W}) + \Delta S^\gamma_h
\ \equiv \ b_\gamma\,S^\gamma_{\rm SM}\,, 
\end{eqnarray}
where $\tau_{x}=M_{H}^2/4m_x^2$, $N_C=3$ for quarks and $N_C=1$ for
taus, respectively.
The additional contribution $\Delta S^\gamma_h$ from non-SM particles
is assumed to be real. Taking $M_H=125.5$ GeV, we find
$S^\gamma_{\rm SM}=-6.64+0.0434\,i$.  
For the loop functions and the normalization of the amplitude, 
we refer to Ref.~\cite{Lee:2003nta}.

The SM Higgs coupling to two gluons is 
given similarly as in $h\to\gamma\gamma$. 
The amplitude for the decay process
$h \rightarrow gg$ can be written as
\begin{eqnarray} \label{higg}
{\cal M}_{gg h}=-\frac{\alpha_s\,M_{H}^2\,\delta^{ab}}{4\pi\,v}\,
S^g_h\,\left(\epsilon^*_{1\perp}\cdot\epsilon^*_{2\perp}\right)
\end{eqnarray}
where $a$ and $b$ ($a,b=1$ to 8) are indices of the eight SU(3)
generators in the adjoint representation.
Again, including some additional loop contributions from new non-SM particles,
the scalar form factor is given by
\begin{eqnarray}
S^g_h=\sum_{f=b,t} b_f\,F_{sf}(\tau_{f}) +  \Delta S^g_h
\ \equiv \ b_g\,S^g_{\rm SM}\,.
\end{eqnarray}
The additional contribution $\Delta S^g_h$ is assumed to be real. 
Taking $M_H=125.5$ GeV, we find
$S^g_{\rm SM}=0.651+0.0501\,i$.  

Finally, for the SM Higgs coupling to $Z$ and $\gamma$,
the amplitude for the decay process $h \to Z(k_1,\epsilon_1)\
\gamma(k_2,\epsilon_2)$ can be written as
\begin{equation}
{\cal M}_{Z\gamma h} = -\,\frac{\alpha}{2\pi v}\,
S^{Z\gamma}_h\,
\left[ k_1\cdot k_2\,\epsilon_1^*\cdot\epsilon_2^*
-k_1\cdot\epsilon_2^*\,k_2\cdot\epsilon_1^* \right] 
\end{equation}
where $k_{1,2}$ are the momenta of the $Z$ boson and the photon (we note that
$2k_1\cdot k_2 = M_{H}^2-M_Z^2$), and
$\epsilon_{1,2}$ are their polarization vectors.
The scalar form factor is given by
\begin{eqnarray}
S^{Z\gamma}_h & \equiv & b_{Z\gamma}\,S^{Z\gamma}_{\rm SM} 
\\
& = &  
2 \sum_{f=t,b,\tau} Q_f N_C^f m_f^2\
\frac{I_3^f-2\, \sin^2 \theta_W Q_f^2}
{\sin\theta_W \cos\theta_W}\, b_f\, F_f^{(0)}
+M_Z^2 \cot\theta_W\, b_W\, F_W 
+ \Delta S^{Z\gamma}_h \
\end{eqnarray}
The additional contribution $\Delta S^{Z\gamma}_h$ is  
 assumed to be real. Taking $M_H=125.5$ GeV, we find
$S^{Z\gamma}_{\rm SM}=-11.0+0.0101\,i$.  
For the loop functions and the normalization of the amplitude, 
we refer to Ref.~\cite{Cheung:2013kla}.

\subsection{Couplings of the singlet scalar and the mixing}
The relative strength of the couplings of the singlet scalar boson $s$
before mixing can be defined similarly in terms of a set of ratios $c_i\;
(i = f,W,Z,\gamma,g)$.  Here the $c_i$ parameterize the couplings of 
$s$ to the SM particles in a way similar to  those of the SM Higgs boson $h$:
\begin{eqnarray}
{\cal L}_{s\bar{f}f}& =& - \sum_{f=u,d,l}\,\frac{g m_f}{2 M_W}\,c_f\,
s\, \bar{f}\,f\,, \\
{\cal L}_{sVV}  &=&  g\,M_W \, \left(
c_W W^+_\mu W^{- \mu}\ + \
c_Z \frac{1}{2\, \cos^2\theta_W}\,Z_\mu Z^\mu\right) \, s\,, \nonumber \\
S^\gamma_s &=& 2\sum_{f=b,t,\tau} N_C\,
Q_f^2\, c_f\,F_{sf}(\tau_{f}) - c_W\,F_1(\tau_{W}) + \Delta S^\gamma_s
\ \equiv \ c_\gamma\,S^\gamma_{\rm SM}\,,  \nonumber \\
S^g_s&=&\sum_{f=b,t} c_f\,F_{sf}(\tau_{f}) +  \Delta S^g_s 
\ \equiv \ c_g\,S^g_{\rm SM}\,, \nonumber \\
S^{Z\gamma}_s & = &
2 \sum_{f=t,b,\tau} Q_f N_C^f m_f^2\
\frac{I_3^f-2\, \sin^2 \theta_W Q_f^2}
{\sin\theta_W \cos\theta_W}\, c_f\, F_f^{(0)}
+M_Z^2 \cot\theta_W\, c_W\, F_W 
+ \Delta S^{Z\gamma}_s \ \equiv c_{Z\gamma}\,S^{Z\gamma}_{\rm SM} \,.
\nonumber
\end{eqnarray}

Since all the relative couplings $c_i$'s come from nonrenormalizable 
interactions between the singlet scalar $s$ and the SM particles, 
except for the Higgs fields, one can simply assume that
$c_i$'s are naturally suppressed by a heavy mass scale or a loop suppression 
factor:
\[
c_i \sim  ``0"  + \frac{g^2 m^2}{(4 \pi)^2 M^2} \,,   
\;\;\; {\rm or}\;\;\;  ``0" + \frac{g^2 m^2}{M^2} \,,
\]
On the other hand, the relative couplings $b_i$'s of the SM Higgs boson 
with deviations coming from higher dimensional operators or additional particles
running in the loop can be expressed as 
\[
b_i \sim ``1" + \frac{g^2 m^2}{(4 \pi)^2 M^2} \,, 
 \;\; {\rm or} \;\;  ``1" + \frac{g^2 m^2}{M^2} \,,
\]
where $M$ is the mass scale of a new particle that has been integrated
out, and $m$ is the external SM particles with $m \ll M$, and $g$ is a
typical coupling of the SM particle and the heavy particle.  Note that
there would be extra loop suppression factors ($\sim 1/(4\pi)^2$) if
the relevant operators are generated at one loop level. 
The sizes of $b_i$'s and $c_i$'s then set the stage for our numerical analysis.

One further complication comes from the mixing between the SM Higgs field $h$
and the singlet field $s$.    The two mass eigenstates $H_{1,2}$
are related to the interaction eigenstates by an $SO(2)$ rotation:
\begin{eqnarray}
H_1  =  h\,\cos\alpha - s\,\sin\alpha\,; \ \ \ \
H_2  =  h\,\sin\alpha + s\,\sin\alpha\,,
\end{eqnarray}
with $\cos\alpha \equiv c_\alpha$ and $\sin\alpha \equiv s_\alpha$ describing 
the mixing between the interaction eigenstates $h$ 
(remnant of the SM Higgs doublet) and $s$ (singlet).   
In this work,  we are taking $H_1\equiv H$ for the 125 GeV boson 
discovered at the LHC and $H_2$ can be either heavier or lighter 
than $H_1$. We are taking $\cos\alpha > 0$ without loss of
generality.

Then, the relative couplings of the observed Higgs boson $H$ 
to fermions $f$, gauge bosons  $W,Z,\gamma,g$ are then given by
\begin{equation}
   b_i \cos\alpha - c_i \sin\alpha \;\;\; (i = f,W,Z,\gamma,g) \;.
\end{equation}

We observe that 
$k_i=(b_i c_\alpha - c_i s_\alpha)^2$, see Eq.~(\ref{eq:kappa1}).
Note that the loop-induced Higgs decay with $i=g,\gamma$ can be
modified by several different origins; (i) from scalar mixing denoted
by $\alpha$, (ii) from the singlet scalar couplings denoted by
$c_{g,\gamma}$, especially when the singlet scalar couples to extra vector-like
quarks and/or leptons or charged vector bosons, (iii) from
modifications of the top and/or $W$ boson couplings in the loop which are
denoted by $b_t$ and $b_W$, which arise from higher dimensional
operators involving the SM Higgs doublet and the SM chiral fermions
with the full SM gauge symmetry, 
(iv) from the couplings of the SM Higgs doublet to extra vector-like
quarks and/or leptons or charged vector bosons,
and (v) from some new physics
effects that directly modify the couplings of the SM Higgs interaction
eigenstate.

The $\kappa_i$ parameterization is effective and simple but is highly
degenerate, since different values of $c_\alpha , b_i , c_i$ can lead
to the same value of $\kappa_i$. It would be impossible to separate 
the true origin of new physics generating $\kappa \neq 1$ in the $\kappa$
parameterization.

\subsection{Signal strength}
%
%
The theoretical signal strengths may be written  as
\begin{equation}
\widehat\mu({\cal P},{\cal D}) \simeq
\widehat\mu({\cal P})\ \widehat\mu({\cal D}) \;,
\end{equation}
where ${\cal P}={\rm ggF}, {\rm VBF}, VH, ttH$ denote the Higgs
production mechanisms: gluon fusion (ggF), vector-boson fusion (VBF),
and associated productions with a $V=W/Z$ boson ($VH$)
and top quarks ($ttH$) 
and ${\cal D}=\gamma\gamma $, $ZZ,$ $WW,$ $b\bar{b},$ $\tau\bar\tau$
the decay channels.

More explicitly, we are taking
\begin{eqnarray}
\widehat\mu({\rm ggF}) &=& (b_gc_\alpha-c_gs_\alpha)^2\,, \ \ \
\widehat\mu({\rm VBF}) = \widehat\mu(VH) = (b_Vc_\alpha-c_Vs_\alpha)^2\,, \ \ \
\widehat\mu(ttH) = (b_tc_\alpha-c_ts_\alpha)^2 \nonumber \\
\end{eqnarray}
with $V=Z,W$ and
\begin{equation}
\widehat\mu({\cal D}) = \frac{B(H\to {\cal D})}{B(H_{\rm SM}\to {\cal D})}
\end{equation}
with
\begin{equation}
B(H\to {\cal D})=\frac{\Gamma(H\to{\cal D})}
{\Gamma_{\rm tot}(H)+\Delta\Gamma_{\rm tot}} 
=\frac{(b_ic_\alpha-c_is_\alpha)^2\,B(H_{\rm SM}\to {\cal D})}
{\Gamma_{\rm tot}(H)/\Gamma_{\rm SM}
+\Delta\Gamma_{\rm tot}/\Gamma_{\rm SM}}\,,
\end{equation}
where $i=\gamma,Z,W,b$ and $\tau$ for 
${\cal D}=\gamma\gamma , ZZ, WW, b\bar{b}$ and $\tau\bar\tau$, respectively.
Note that we introduce an arbitrary non-SM contribution $\Delta\Gamma_{\rm tot}$
to the total decay width. Incidentally,
$\Gamma_{\rm tot}(H)$ becomes the SM total decay width $\Gamma_{\rm SM}$
when $c_\alpha=1$, $b_f=b_V=1$, 
$\Delta S^{\gamma,g,Z\gamma}_h=0$
\footnote{We note $b_{\gamma,g,Z\gamma}=1$ when $b_f=b_V=1$ and
$\Delta S^{\gamma,g,Z\gamma}_h=0$.}, 
and $\Delta \Gamma_{\rm tot} = 0$.
For more details, we refer to Ref.~\cite{Cheung:2013kla}.

\section{Models}
In a number of phenomenologically well motivated BSM models, there often
appears a SM singlet scalar boson that can mix with the SM Higgs boson.
Adding an extra singlet field to the SM is the simplest extension of 
the SM Higgs sector in terms of new degrees of freedom.
A singlet scalar boson $s$ does not affect the $\rho$ parameter at tree level, 
and
is not that strongly constrained by the electroweak precision tests (EWPT).
It can also make the electroweak phase transition strongly first 
order~\cite{1st_ewpt},
and enables us to consider electroweak baryogenesis if there are new sources
of CP violation beyond the Kobayashi-Maskawa (KM) phase in the  SM with three
generations.  
Finally,  if we imposed a new discrete $Z_2$  symmetry
$s\rightarrow -s$, the singlet scalar  $s$ could make a good dark matter
candidate~\cite{singlet_scalar_dm}.   
This is the standard list for the rationales for considering
a singlet scalar $s$.

However, there are many more interesting scenarios where a singlet scalar
appears in a natural way and plays many important roles.
Let us list some examples, referring to Ref.~\cite{Chpoi:2013wga}
for more extensive discussion.

\subsection{Dark matter models  with dark gauge symmetries and/or Higgs portals}
First of all, let us consider DM models where weak scale DM is stabilized by
some spontaneously broken local dark gauge symmetries~\cite{Hur:2007uz,
Hur:2011sv,Baek:2012se,Baek:2013dwa,Baek:2013qwa,Baek:2014kna,Ko:2014bka,
Ko:2014gha,Ko:2014loa,Ko:2014lsa,Ko:2014nha,Ko:2015ioa}. This possibility is not
that often considered seriously. However if we remind ourselves of the
logic behind $U(1)_{\rm em}$ gauge invariance, electric charge conservation,
existence of massless photon and electron stability and non-observation of
$e \rightarrow \nu \gamma$, one would realize immediately the same logic could
be applied to the DM model building. One might think that this assumption
may be too strong, since the lower bound on the DM lifetime is much weaker 
than   that on the proton lifetime.  
This is in fact true, but this can be understood since 
proton is a composite particle, a bound state of 3 quarks with 
color gauge interaction, 
and baryon number violating operator in the SM is dim-6 
or higher.  Likewise longevity of DM
might be due to some new strong interactions that make 
DM particle composite.
Also, considering all the SM particles feel some gauge interactions,
it would be natural to assume that the DM also may feel some gauge interactions
(see Ref.~\cite{Ko:2015vaa} for a recent review).

In the case the dark matter particle is associated with 
some dark gauge symmetries,
there would  generically appear a dark Higgs boson after dark gauge symmetry
breaking. The original dark Higgs $\Phi$ would be charged under some local dark
gauge symmetry, but it is a singlet under the SM gauge group 
in the simplest setup.
And after dark gauge symmetry breaking, there would be dark Higgs boson 
$h_\Phi$,
which would mix with the SM Higgs boson via the Higgs-portal interaction,
\[
\lambda_{H\Phi} ( H^\dagger H - \frac{v^2}{2} )( \Phi^\dagger 
\Phi - \frac{v_\Phi^2}{2} ) \;.
\]
A Higgs-portal coupling as small as $\lambda_{H\Phi} \sim 10^{-6}$ can
thermalize the hidden sector DM efficiently
\footnote{See, for example, Sec. III E and Fig.~5 
(right panel) in Ref.~\cite{rp4} for more details.}.
On the other hand, the effects of 
such a small coupling would be very difficult to observe at colliders.

Also, the dark Higgs can stabilize the EW vacuum up to Planck 
scale, as well as it
can modify the standard Higgs inflation scenario in such a way that a large
tensor-to-scalar ratio $r \sim (0.1)$ could be
 possible~\cite{Ko:2014eia}, which is 
independent of the precise values for the top quark and/or Higgs boson masses.
Although the dark Higgs boson was introduced in order to break the 
dark gauge symmetry spontaneously, it has additional niceties 
in regard of cosmology in the context of
the EW vacuum stability and the Higgs inflation assisted by 
Higgs-portal interaction.

Even if we relax the assumption of the local dark gauge symmetry and consider
more phenomenological Higgs-portal DM models, there will still appear a singlet
scalar boson that can mix with the SM Higgs boson,
if the Higgs-portal DM is a singlet Dirac fermion  ~\cite{Baek:2012uj,Baek:2011aa} or a 
vector boson~\cite{Baek:2012se,Farzan:2012hh,Duch:2015jta}.  
Also, it can play an important role in DM phenomenology. 
For example, one can easily accommodate the galactic center $\gamma$-ray excess 
by DM pair annihilation into a pair of dark Higgs bosons, followed by 
dark Higgs decays into the SM 
particles~\cite{Boehm:2014bia,Ko:2014gha,Ko:2014loa,Ko:2014nha,Ko:2015ioa}.

Furthermore, there could be non-standard Higgs decays into a pair of lighter neutral
scalar bosons (namely the dark Higgs boson) or a pair of dark gauge bosons, 
in addition 
to a pair of dark matter particles. In this case, the total decay width of 
the observed
Higgs boson would receive additional contributions from the final states with
dark matter, dark gauge bosons, or dark Higgs bosons, which are parameterized
in terms of $\Delta \Gamma_{\rm tot}$.
Therefore, we will take the deviation $\Delta \Gamma_{\rm tot}$ in the total decay 
width of the 125 GeV Higgs boson 
as a free parameter when we perform global fits
to the LHC data on the Higgs signal strengths.

These classes of BSM models are phenomenologically very well motivated,
and they have very  significant  impacts on the observed 125 GeV scalar boson.
Therefore, it is very important to seek for a singlet scalar that can mix with
the SM Higgs boson in all possible ways. The phenomenology 
associated with the observed Higgs boson measurements
is straightforward. The signal strengths of the 125GeV Higgs boson
are suppressed from ``1'' in a universal manner, namely independent of 
production and decay channels. Moreover,
the 125 GeV Higgs couplings to SM fermions and weak gauge bosons are all
suppressed by $\cos\alpha$ relative to the SM values.
One can also search for the heavier Higgs boson in this type of Higgs-portal 
models~\cite{collider}.

In summary, hidden-sector DM models are characterized by 
$b_i = 1$ and $c_i=0$ with a few simple implications:
\begin{itemize}
\item Couplings to the SM fermions and gauge bosons are all suppressed by the
factor $\cos\alpha$.
\item Decay Width: 
$\Gamma({H\to {\cal D}})=\cos^2\alpha\, \Gamma_{\rm SM}({H\to {\cal D}})$
and $\Gamma_{\rm tot}(H)=\cos^2\alpha\,\Gamma_{\rm SM}$.
Note that the total decay width of the Higgs boson, including the
non-SM decay modes, is given by
$ \Gamma_{\rm tot} (H) + \Delta \Gamma_{\rm tot}$.
\item Signal strengths:
$\widehat\mu({\cal P},{\cal D}) \simeq
\widehat\mu({\cal P})\ \widehat\mu({\cal D}) =
\frac{\cos^4\alpha}{\cos^2 \alpha+\Delta\Gamma_{\rm tot}/\Gamma_{\rm SM}}$
independently of the production mechanism ${\cal P}$ and 
the decay channel ${\cal D}$.
\item Varying parameters:
$\cos\alpha$ and $\Delta \Gamma_{\rm tot}$.
\end{itemize}
In terms of two free parameters $\cos\alpha$ and $\Delta \Gamma_{\rm tot}$,
we perform the $\chi^2$ minimization procedures on the LHC Higgs signal strength
data in the next section. 

\subsection{Non-SUSY $U(1)_{B-L}$ extensions of the SM}

Another interesting example of Higgs-portal models is the  nonsupersymmetric 
$U(1)_{B-L}$ extension of the  SM plus 3 RH neutrinos, which is anomaly free,
so that no new colored or EW charged fermions are introduced:
\begin{equation} 
{\cal L}= {\cal L}_{\rm SM} - V(H,\Phi)  -  
\left(\frac{1}{2}\lambda_{N,i}\Phi \bar N_i^c N_i+Y_{N,ij}
\bar \ell H^\dagger N+h.c.\right) \;,
\end{equation}
where the scalar potential $V(H,\Phi)$ is given by  
\begin{equation} 
V(H,\Phi)=  - \mu_H^2 H^\dagger H - \mu_\phi^2 \Phi^\dagger \Phi  
+ \frac{\lambda_{h}}{2}|H|^4 -\lambda_{h\phi}|H|^2|\Phi|^2+
\frac{ \lambda_\phi}{2}|\Phi|^4 \,.
\end{equation}
Here the SM singlet scalar $\Phi$ carries $B-L$ charge ``2'', 
and after $B-L$ symmetry breaking from 
the nonzero VEV of $\Phi$,   the resulting singlet scalar $\phi$  
will mix with  the SM Higgs field.

If the $B-L$ gauge boson $Z^{'}$ is light enough, the observed Higgs boson 
can decay into a pair of $Z^{'}$ bosons through the mixing between the SM
Higgs boson and the $U(1)_{B-L}$-charged singlet scalar $\phi_{B-L}$,
if this decay is kinematically allowed.  The current bound on this
model from the Drell-Yan process is that $v_\phi \gtrsim$ a few TeV, so
that $g_{B-L} \sim$ (a few) $\times 10^{-3}$ or less, for this to
happen.  
In this case, the Higgs phenomenology is described by two 
parameters, $\cos\alpha$ and $\Delta \Gamma_{\rm tot}$, as in DM models
with dark gauge symmetry and/or Higgs portals.

\subsection{Vector-like fermions for enhanced 
$H(125) \rightarrow \gamma\gamma$ and/or
$H(125) \rightarrow gg$}

Right after the first candidate signature for the SM Higgs boson 
at the LHC, a
number of groups considered new vector-like fermions (quarks or
leptons) in order to explain the excessive signal strength in the
$H\rightarrow \gamma \gamma$ decay channel.  When considering
vector-like fermions, one often has to introduce a singlet scalar field at
renormalizable interaction level. 
Note that vector-like fermions
can not directly couple to the SM Higgs doublet, 
and one has to introduce a singlet  scalar coupled to them.
Only in the presence of a singlet  scalar, therefore,
they can couple to the SM Higgs through the mixing 
between the SM Higgs boson and the new singlet scalar
\footnote{More detailed discussions  of this class of models can be found in 
Sec. 3.4 in Ref.~\cite{Chpoi:2013wga}.}.
In this case,
the mixing between the SM Higgs boson and the new singlet scalar
tends to reduce the signal strength for $H\rightarrow \gamma \gamma$
decay channel, although the loop contributions from the vector-like
fermions would generate the singlet scalar decays into $\gamma\gamma$
and/or $gg$.  It is essential to consider the mixing effects in the
proper way (see, for example, Ref.~\cite{Abe:2012fb}).

In these types of models, the observed Higgs-boson couplings are given by
\begin{eqnarray}
g^S_{H\bar{f}f} &= & (b_f\cos\alpha-c_f\sin\alpha) =\cos\alpha \,; \nonumber \\[2mm]
g^S_{HVV} &= & (b_V\cos\alpha-c_V\sin\alpha) =\cos\alpha \ \ \
{\rm for} \ \ \ V=Z,W\,; \nonumber \\[2mm]
S^{\gamma,g,Z\gamma}_H  &= & 
(S^{\gamma,g,Z\gamma}_h\cos\alpha-S^{\gamma,g,Z\gamma}_s\sin\alpha) \nonumber \\
&=&\cos\alpha\, S^{\gamma,g,Z\gamma}_{\rm SM}
+ \left(\Delta S^{\gamma,g,Z\gamma}_h\, \cos\alpha
- \Delta S^{\gamma,g,Z\gamma}_s\, \sin\alpha\right) \nonumber \\
&\equiv& \cos\alpha\, S^{\gamma,g,Z\gamma}_{\rm SM}+\Delta S^{\gamma,g,Z\gamma}_H\,.
\end{eqnarray}
assuming $b_f = b_V= 1$ and  $c_f = c_V = 0$.

In this case, the varying parameters are $\cos\alpha$, 
$\Delta S^\gamma_{h,s}$ and/or $\Delta S^g_{h,s}$, 
and possibly including 
$\Delta\Gamma_{\rm tot}$.

\subsection{Summary of the models}

Here we summarize the models in which a singlet scalar boson
mixes with the SM Higgs boson: see Table \ref{tab:cici} for the relevant $c_F$'s.
More details including the corresponding Lagrangian for each model 
can be found in Ref.~\cite{Chpoi:2013wga}.

Note that those classes of BSMs described in the subsections
III.A and III.B are phenomenologically very well motivated by dark
matter and neutrino physics as well as grand unification.  Also, their
impacts on the observed 125 GeV scalar boson as well as on the
EW vacuum stability or Higgs inflation are straightforward:
\begin{itemize}
\item The signal strengths of the 125 GeV Higgs boson are suppressed
from ``1'' in a universal manner,
namely independent of production and decay channels.
\item The 125 GeV Higgs couplings to the SM fermions and
the weak gauge bosons are all suppressed by $\cos\alpha$
relative to the SM values.
\item The additional singlet scalar boson can improve the stability of
EW vacuum up to the  Planck scale~\cite{Baek:2012uj}. 
\item The singlet scalar can improve the EW phase transition 
to be more strongly first order.
\item The mixing between the SM Higgs boson and the singlet scalar boson
can modify the predictions for the tensor-to-scalar ratio within the
 Higgs inflation, 
and disconnecting  the strong correlation of the inflationary
observables from the top quark and the Higgs boson masses.
\end{itemize}
Therefore it is very important to seek for a singlet scalar boson
that can mix with the SM Higgs boson in all possible ways.

\begin{table}[t]
\caption{\small \label{tab:cici} 
Nonvanishing $c_F$'s in various BSMs with an extra singlet
  scalar boson.  For non-SUSY $U(1)_{B-L}$ model, there would be
  nonzero $c_{Z^{'}}$ where $Z^{'}$ is the $U(1)_{B-L}$ gauge
  boson. Since the bound from Drell-Yan is very stringent, we will
  ignore $H (125) \rightarrow Z^{'} Z^{'}$, although it is in
  principle possible if the gauge coupling is very small $g_{B-L}
  \lesssim$ a few $\times 10^{-3}$. 
  Details can be found in Ref.~\cite{Chpoi:2013wga}.
 }
\begin{center}
\begin{tabular}{c|c}
\hline \hline
Model   &     Nonzero  $c_F$'s     
\\
\hline \hline
Pure Singlet Extension &   $c_{h^2} $ 
\\
Hidden Sector DM &   $c_\chi $,$c_{h^2} $ 
\\
non-SUSY $U(1)_{B-L}$ & $c_{Z^{'}}$, $c_{h^2}$  
\\
Dilaton & $c_g ,  c_W , c_Z , c_\gamma ,c_{h^2}$
\\
Vector-like Quarks &   $ c_g , c_\gamma , c_{Z\gamma} , c_{h^2} $ 
\\
Vector-like Leptons &  $c_\gamma , c_{Z\gamma}, c_{h^2} $  
\\
New Charged Vector bosons &   $c_\gamma , c_{h^2} $ 
\\
Extra charged scalar bosons &  $c_g , c_\gamma , c_{Z\gamma}, c_{h^2}$ 
\\
\hline \hline
\end{tabular}
\end{center}
\end{table}

\section{Results}
%
We are going to perform the following fits:
\begin{itemize}
\item {\bf SD} fit -- Singlet Dark Matter model and non-SUSY $U(1)_{B-L}$ case: 
Varying $c_\alpha$ and $\Delta\Gamma_{\rm tot}$,
\item {\bf SL} fit -- Singlet plus a vector-like Lepton:
Varying $s_\alpha$, $\Delta\Gamma_{\rm tot}$,
$\Delta S^\gamma_h$, and $\Delta S^\gamma_s$,
\item {\bf SQ} fit  -- Singlet plus a vector-like Quark: 
Varying $s_\alpha$, $\Delta\Gamma_{\rm tot}$,
$\Delta S^\gamma_h$, $\Delta S^\gamma_s$,
$\Delta S^g_h$, and $\Delta S^g_s$.
\end{itemize}
Note, instead of $c_\alpha$ 
we vary $s_\alpha$ in the {\bf SL} and {\bf SQ} fits because we have to
specify $c_\alpha$ and $s_\alpha$ simultaneously in these fits.
Otherwise, one may possibly explore the unphysical regions of 
$\Delta\Gamma_{\rm tot}<0$ and
$c_\alpha>1$ in the {\bf SD} fit in order to study the parametric dependence.
We neglect the $S^{Z\gamma}_{h,s}$ couplings 
since we do not have any predictive power in
the model-independent approach taken in this work.

We use the most updated data summarized in Ref.~\cite{Cheung:2014noa} and
the results of the fits are summarized in Table~\ref{tab:fits}. We find that
the best-fit values of the {\bf SD} fit are extremely close to the SM ones.
For the {\bf SL} and {\bf SQ} fits, we observe that the best-fit values for 
$\Delta S^{\gamma,g}_h$ and $\Delta S^{\gamma,g}_s$
are large while those for $\Delta S^\gamma_H$ and $\Delta S^g_H$ 
are only about $-0.8$ and $0.02$, respectively.
In the remaining part of this Section, we discuss the details of each fit.

%
\begin{table}[t!]
  \caption{\small \label{tab:fits}
The best-fitted values for {\bf SD}, {\bf SL}, and {\bf SQ} fits.
The SM chi-square per degree of freedom is $\chi^2_{\rm SM}$/d.o.f.$=16.76/29$,
and $p$-value$=0.966$.
  }
\begin{ruledtabular}
\begin{tabular}{ l|ccc|rcrrrrrrr}
Fits  & $\chi^2$ & $\chi^2$/dof & $p$-value &
\multicolumn{9}{c}{Best-fit values} \\
& & & & $s_{\alpha}$ & $\Delta{\Gamma}_{tot}\;[\rm{MeV}]$ &
$\Delta S^\gamma_h$ & $\Delta S^\gamma_s$ & 
$\Delta S^g_h$ & $\Delta S^g_s$ &
$c_{\alpha}$ & $\Delta S^{\gamma}_H$ & 
$\Delta S^{g}_H$ \\
\hline\hline
{\bf SD} & $16.76$ & $0.621$ & $0.937$ & $0.000$ &
$0.000$ & $-$ & $-$ & $-$ & $-$ & $1.000$ & $-$ & $-$  \\
\hline
{\bf SL}
& $15.66$ & $0.626$ & $0.925$ & $0.129$ &
$0.137$ & $-2.953$ & $-16.27$ & $-$ & $-$ & $0.992$ & $-0.835$ & $-$ \\
\hline
{\bf SQ}
& $15.59$ & $0.678$ & $0.872$ & $0.036$ &
$0.357$ & $0.875$ & $46.84$ & $1.315$ & $35.54$ & $0.999$ & $-0.832$ & $0.019$ \\

\end{tabular}
\end{ruledtabular}
\end{table}

\subsection {\bf SD}
In the {\bf SD} fit,
we scan the regions of parameters:
$c_{\alpha}\subset [0:2]$, $\Delta{\Gamma}_{\rm tot}\subset [-4:8\; \rm{MeV}]$
including unphysical regions of $\Delta\Gamma_{\rm tot}<0$ and
$c_\alpha>1$ to study the parametric dependence.

\begin{figure}[t!]
\centering
\includegraphics[height=3.0in,angle=-90]{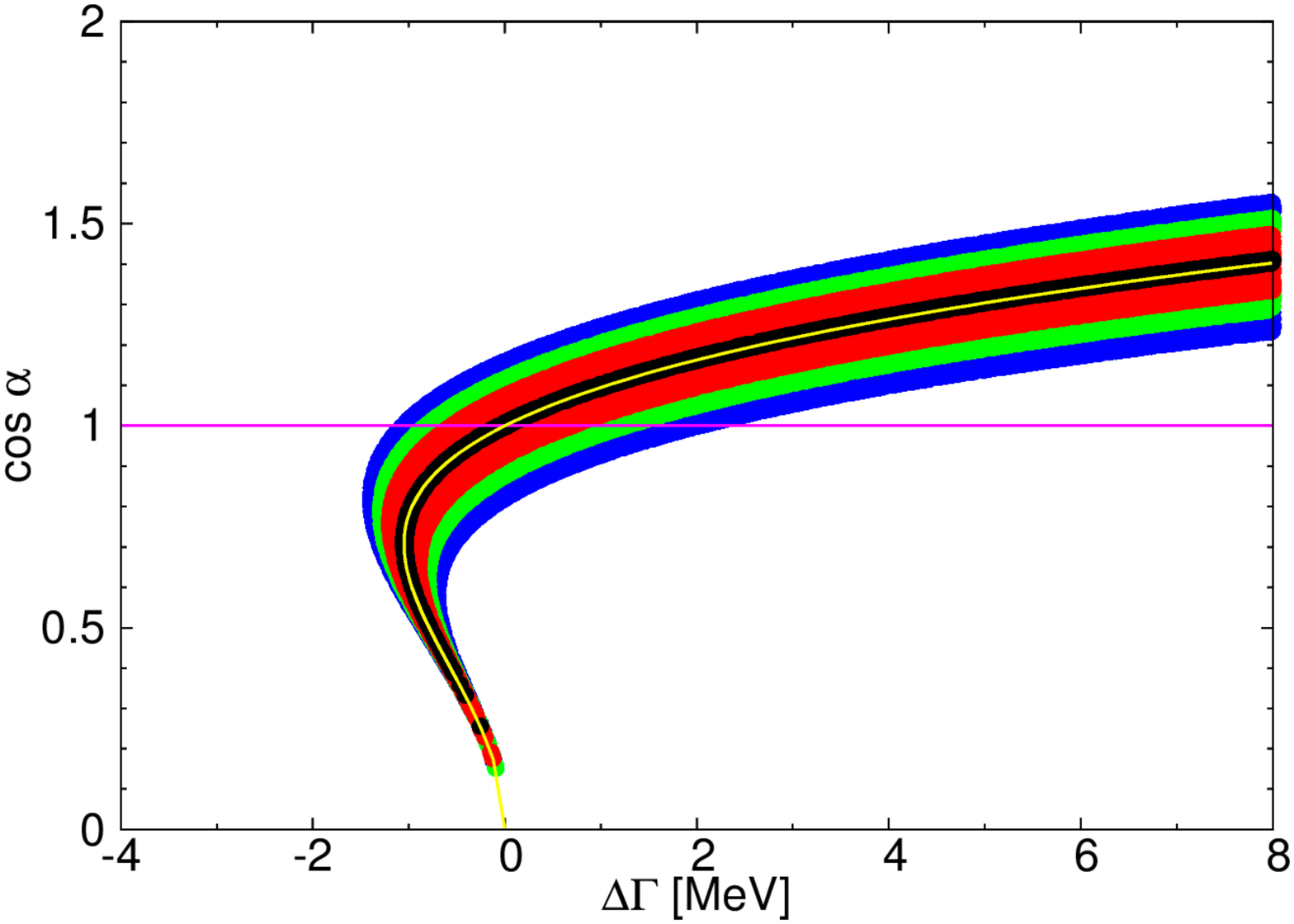}
\includegraphics[height=3.0in,angle=-90]{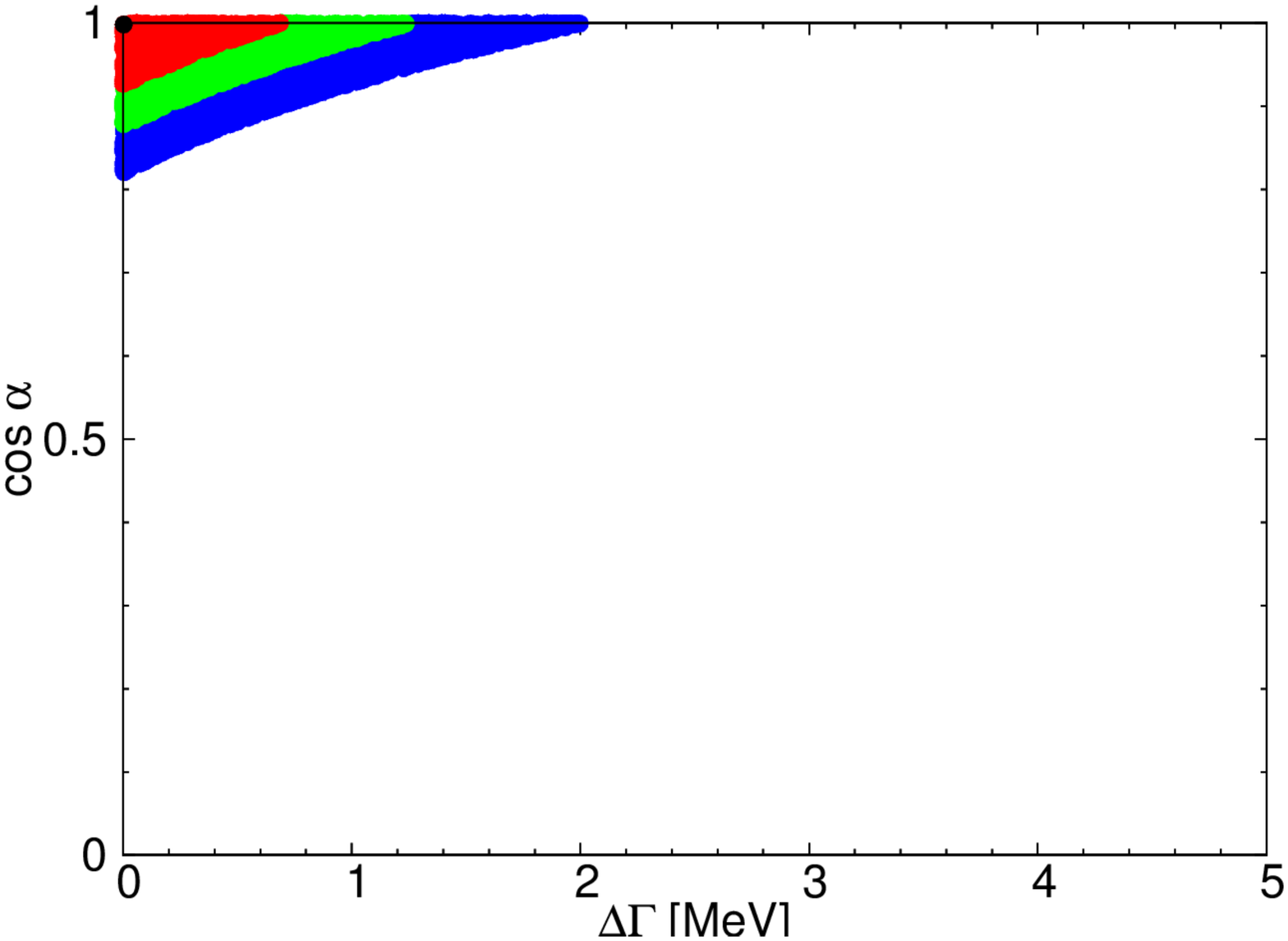}
\caption{\small \label{fig:sd}
The 68 \% ($\Delta\chi^2=2.30$), 95 \% ($\Delta\chi^2=6.18$),
99.7 \% ($\Delta\chi^2=11.83$) confidence-level (CL) regions  
for the {\bf SD} fit
on the $\Delta\Gamma_{\rm tot}$-$\cos\alpha$ plane.
The horizontal line in the left frame shows the physical limit $\cos\alpha=1$.
In the right frame, we show the CL regions after applying
$\Delta\Gamma_{\rm tot}\geq 0$ and $\cos\alpha\leq 1$.
The best-fit points are
along the yellow line passing through the point
$(\Delta\Gamma,\cos\alpha)=(0,1)$ in the left frame.
In the black regions, we have $\Delta\chi^2<0.01$.
}
\end{figure}

In the left frame of Fig.~\ref{fig:sd}, we show
the 68 \% ($\Delta\chi^2=2.30$), 95 \% ($\Delta\chi^2=6.18$),
99.7 \% ($\Delta\chi^2=11.83$) regions on the
$\Delta\Gamma_{\rm tot}$-$\cos\alpha$ plane. 
When $\Delta\Gamma_{\rm tot}\geq 0$, we observe that the minima are developed
along the yellow line in the black ($\Delta\chi^2<0.01$) region which is given by
the relation
\begin{equation}
\label{eq:sm_limit}
\cos\alpha =\left[\frac{1}{2}\left(
1+\sqrt{1+4\frac{\Delta\Gamma_{\rm tot}}{\Gamma_{\rm SM}}}
\right) \right]^{1/2}\, \geq \, 1\,.
\end{equation}
In fact,
the above relation can be obtained by requiring each signal strength to 
be the same as the SM one or
$\widehat\mu({\cal P},{\cal D})=c_\alpha^4/(c_\alpha^2+\Delta\Gamma_{\rm tot}/\Gamma_{\rm
SM})=1$.  This implies the 
the best $\chi^2$ is obtained in the unphysical region of $c_\alpha >1$. When
$c_\alpha \leq 1$, we observe that the best $\chi^2$ is obtained 
again in the unphysical region of $\Delta \Gamma_{\rm tot} < 0$. 
If this is still the case in the future data, a large class of DM
models (Higgs-portal fermion or vector DM, and DM models with 
local dark gauge symmetries) will be disfavored  compared to the SM, 
except for the Higgs-portal scalar DM model without extra singlet scalar, 
for which the Higgs signal strength will be the same as the SM case.
From the right frame of Fig.~\ref{fig:sd}, we see
$\cos\alpha\gsim 0.86\,(0.81)$ and
$\Delta\Gamma_{\rm tot}\lsim 1.24\,(2)$ MeV at $95\%$ ($99.7\%$) CL.
\subsection {\bf SL}

In the {\bf SL} fit,
we scan the regions of parameters:
$s_{\alpha}\subset [-1:1]$, $\Delta{\Gamma}_{\rm tot}\subset [0:4\; \rm{MeV}]$,
$\Delta S^{\gamma}_h\subset [-10:10]$, $\Delta S^{\gamma}_s\subset [-100:100]$.

\begin{figure}[t!]
\centering
\includegraphics[height=2.0in,angle=-90]{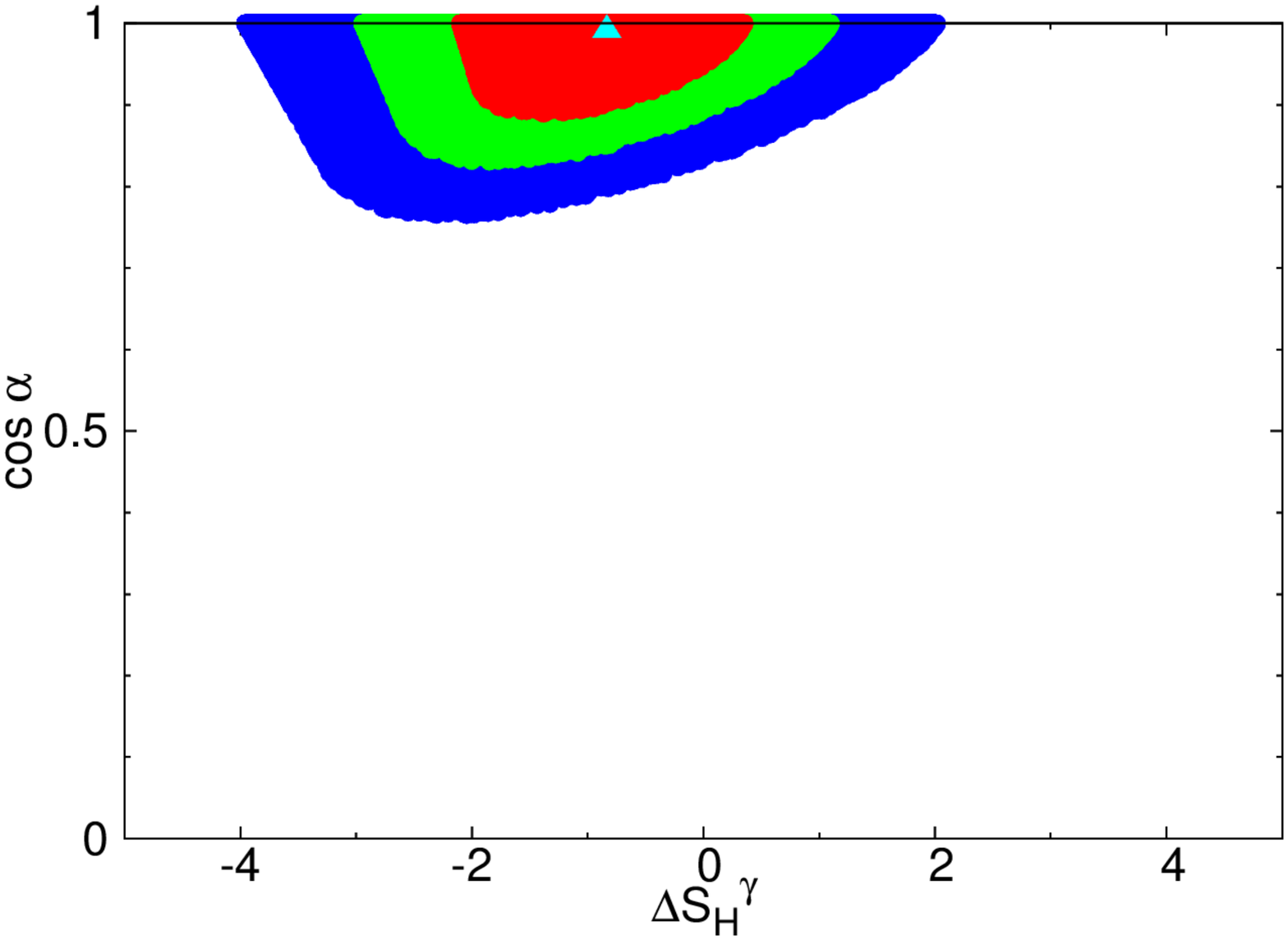}
\includegraphics[height=2.0in,angle=-90]{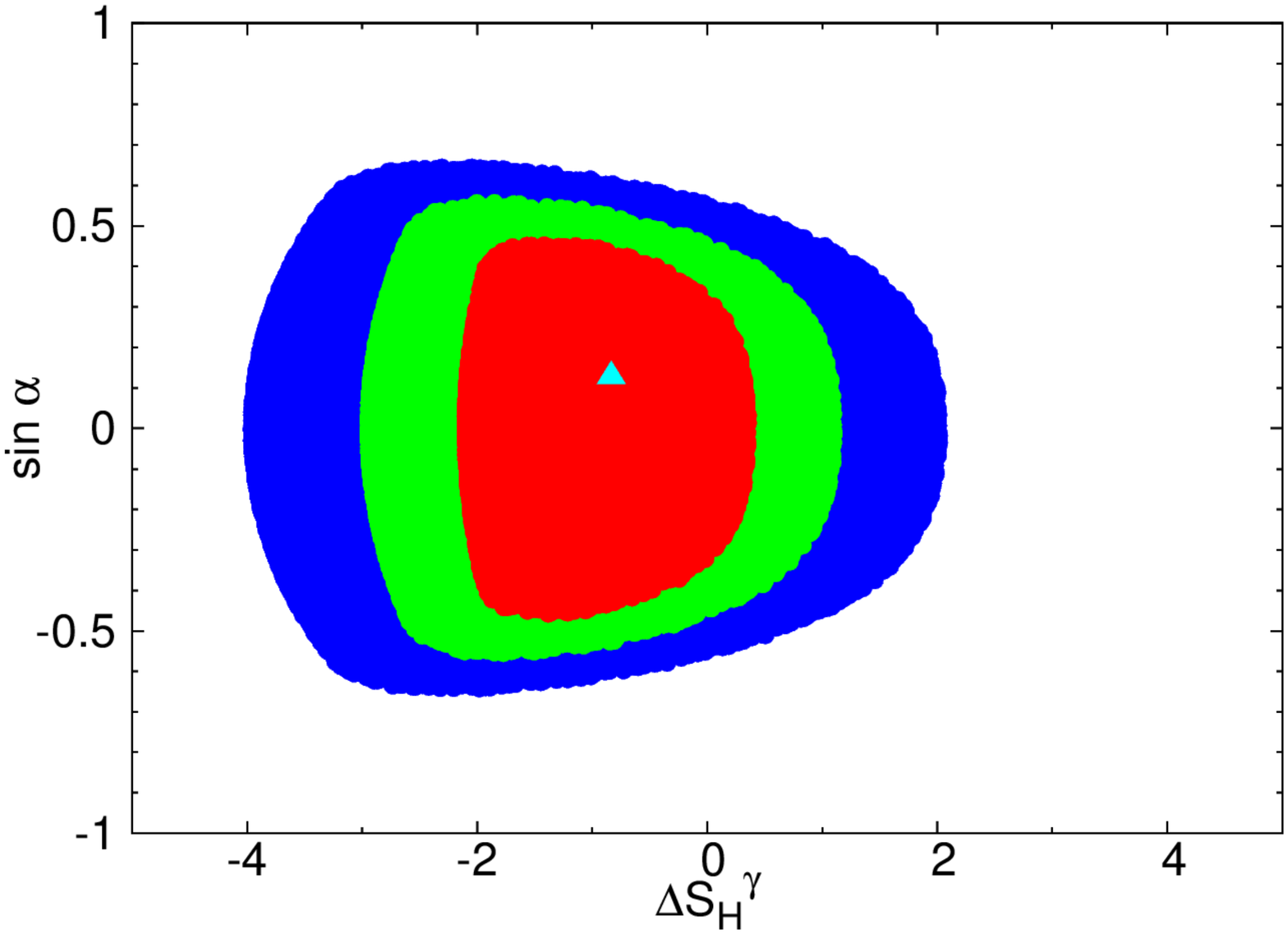}
\includegraphics[height=2.0in,angle=-90]{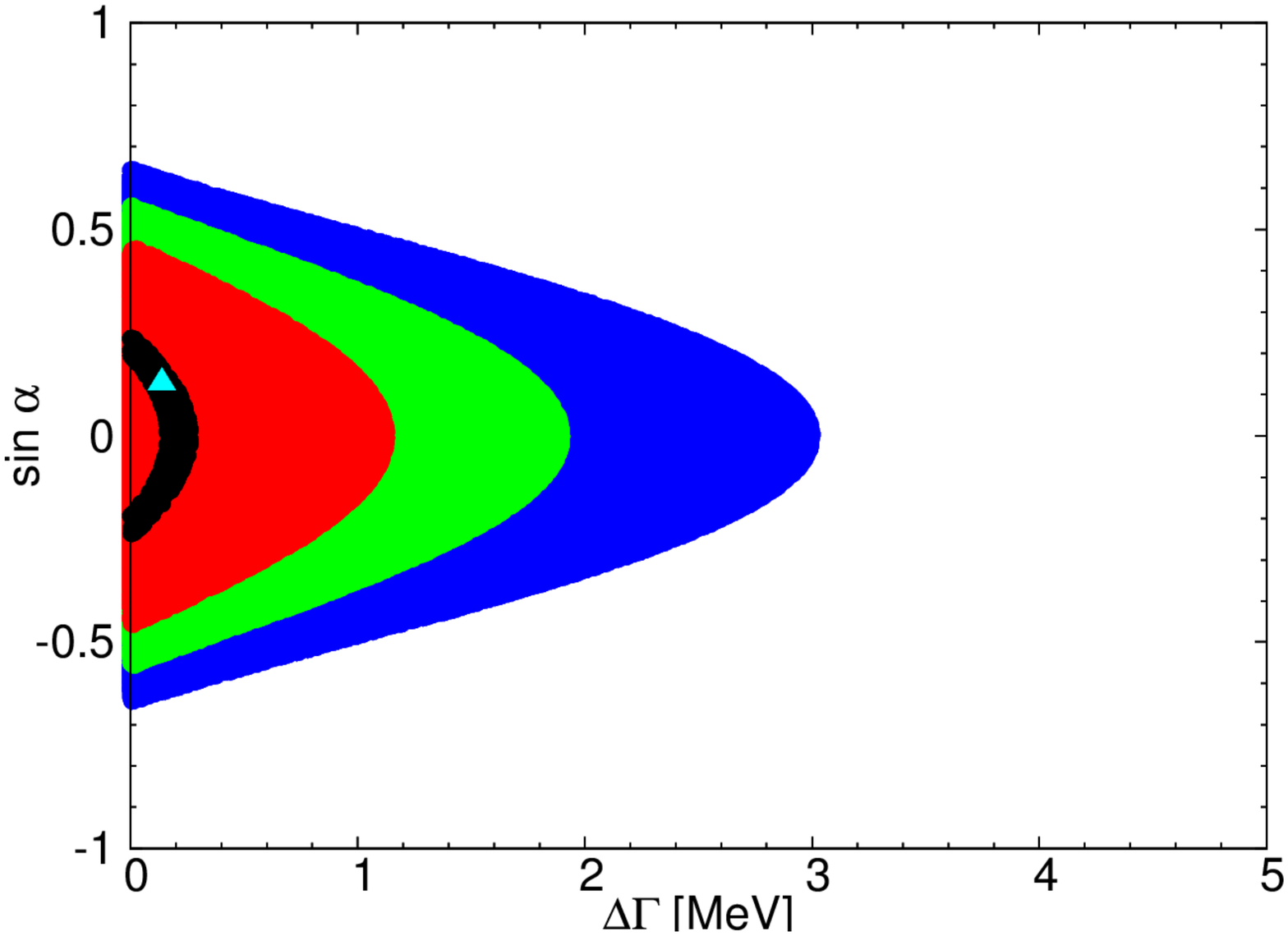}
\includegraphics[height=2.0in,angle=-90]{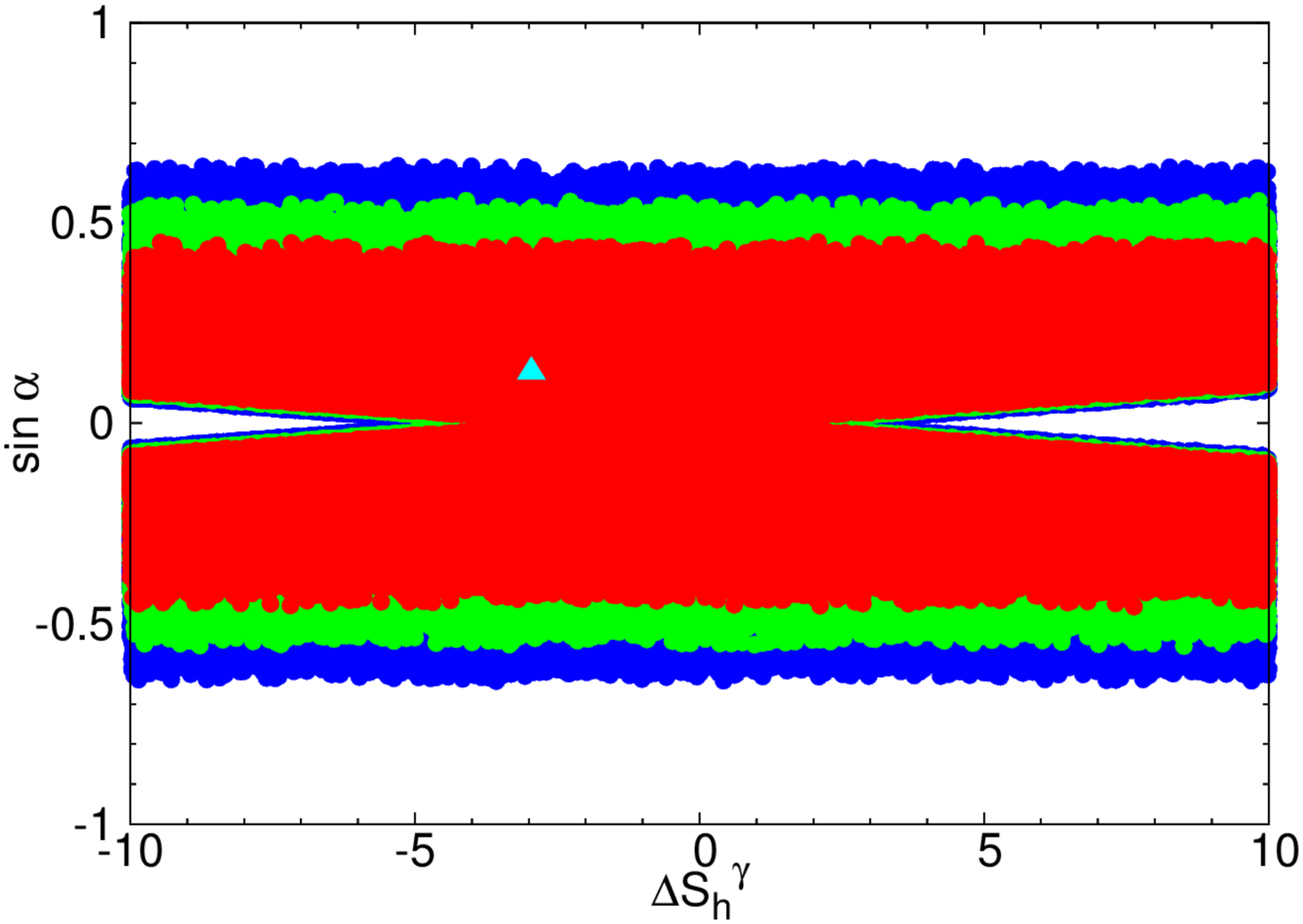}
\includegraphics[height=2.0in,angle=-90]{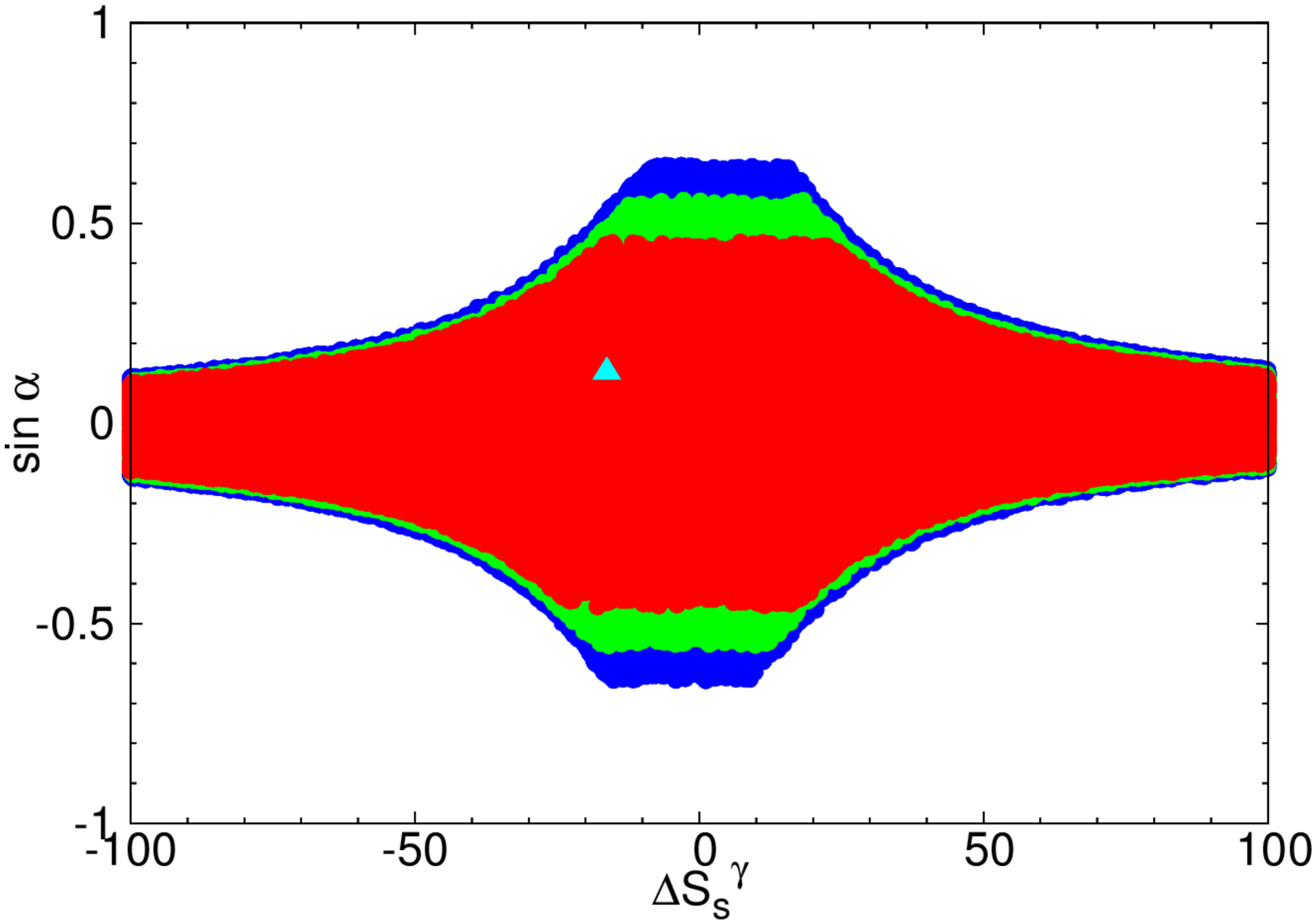}
\includegraphics[height=2.0in,angle=-90]{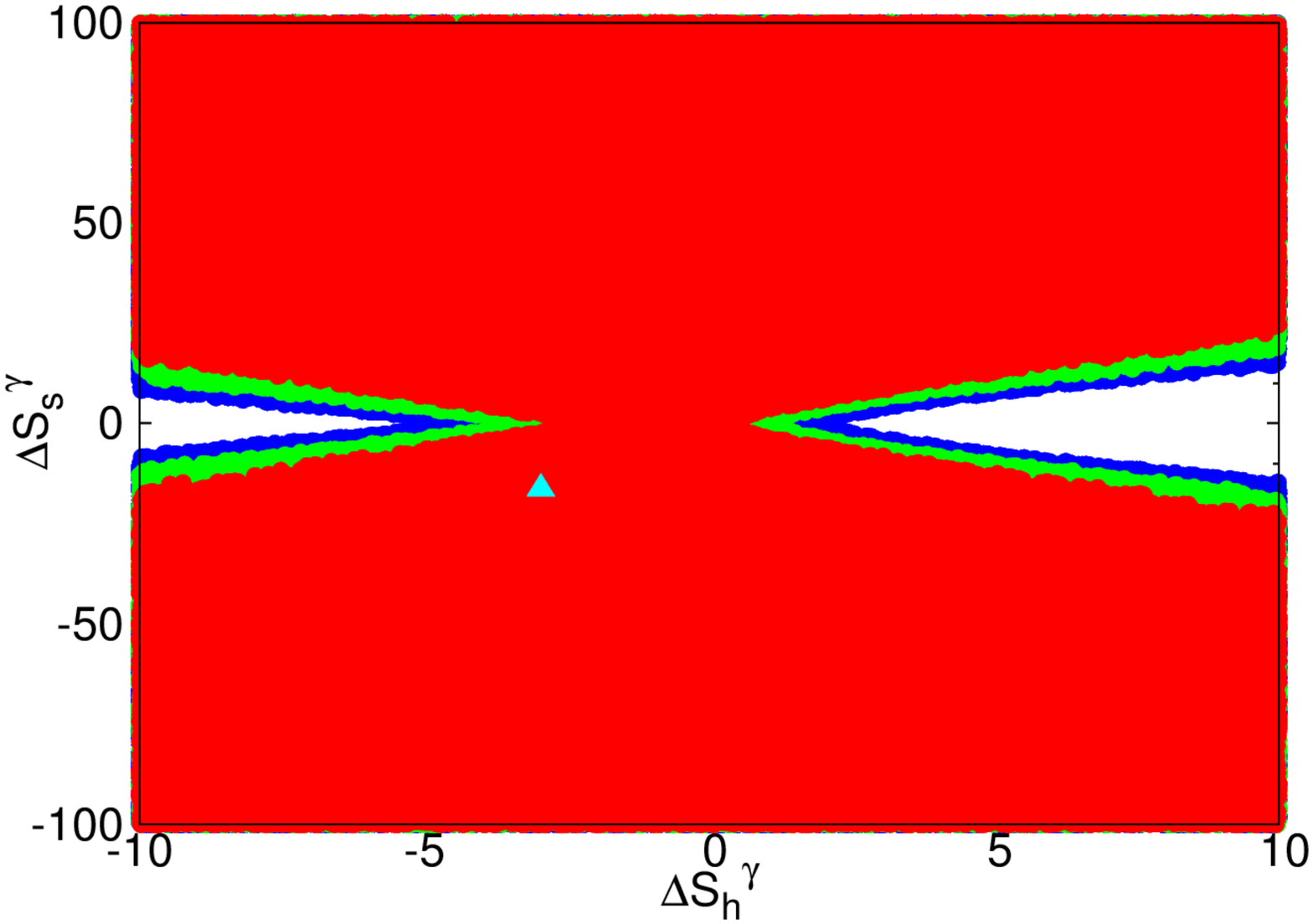}
\caption{\small \label{fig:sl}
The CL regions  for the {\bf SL} fit.
The description of the CL regions is the same as in Fig.~\ref{fig:sd}.
}
\end{figure}

The CL regions are shown in Fig.~\ref{fig:sl}.
We observe that $\cos\alpha$ , $\sin\alpha$, $\Delta\Gamma_{\rm tot}$
and  $\Delta S^\gamma_H$ are well bounded as:
\begin{eqnarray}
&& \cos\alpha \gsim 0.83\,(0.76) \ {\rm at} \ 
95\%\, (99.7\%) \ {\rm CL}\,; \nonumber  \\[2mm]
&& |\sin\alpha| \lsim 0.56\,(0.65)\, \ {\rm at} \ 
95\%\, (99.7\%) \ {\rm CL}\,; \nonumber  \\[2mm]
&& \Delta\Gamma_{\rm tot} \lsim 1.90\,(3.00)\,{\rm MeV} \ {\rm at} \ 
95\%\, (99.7\%) \ {\rm CL}\,; \nonumber  \\[2mm]
&& -2.95\,(-3.96) \lsim \Delta S^\gamma_H \lsim 1.10\,(2.02)\, \ {\rm at} \ 
95\%\, (99.7\%) \ {\rm CL}\,.
\end{eqnarray}
In contrast, $\Delta S^\gamma_h$ and $\Delta S^\gamma_s$ 
are not bounded. From the relation
$\Delta S^\gamma_H= \Delta S^\gamma_h\, \cos\alpha - \Delta S^\gamma_s\, \sin\alpha$,
in the limit $\sin\alpha=0$, we see that 
$\Delta S^\gamma_h=\Delta S^\gamma_H$ is bounded
while $\Delta S^\gamma_s$ can take on any values.
When $|\sin\alpha|$ takes its largest value,
$\Delta S^\gamma_s$ is most bounded:
$|\Delta S^\gamma_s|\lsim 20$, see the lower-middle frame of Fig.~\ref{fig:sl}.
As $|\Delta S^\gamma_h|$ grows, a cancellation between 
the two terms $\Delta S^\gamma_h\, \cos\alpha$
and $\Delta S^\gamma_s\, \sin\alpha$ is needed to obtain the limited
value of $\Delta S^\gamma_H$ 
together with non-vanishing $\Delta S^\gamma_s\, \sin\alpha$, 
explaining the wedges in the lower-left and lower-right frames.

The best-fit values for $\Delta S^\gamma_h$ and $\Delta S^\gamma_s$
are $-2.953$ and $-16.27$, respectively, even though 
$\Delta\chi^2$ does not change much
in most regions of the parameter space. Considering
$S^\gamma_{\rm SM}=-6.64$ and the best-fit value $\Delta S^\gamma_H=-0.835$,
let alone a certain level of cancellation,   it would be very hard to achieve such 
large values for $\Delta S^\gamma_h$  and $\Delta S^\gamma_s$,  unless
the vector-like leptons are light, come in with a large multiplicity,
and/or their Yukawa couplings to the singlet scalar $s$ are strong.

\subsection {\bf SQ}

In the {\bf SQ} fit,
we scan the regions of parameters:
$s_{\alpha}\subset [-1:1]$, $\Delta{\Gamma}_{\rm tot}\subset [0:15\; \rm{MeV}]$,
$\Delta S^{\gamma}_h(\Delta S^{g}_h)\subset [-10:10]$, 
$\Delta S^{\gamma}_s(\Delta S^{g}_s)\subset [-100:100]$. 

\begin{figure}[t!]
\centering
\includegraphics[height=2.0in,angle=-90]{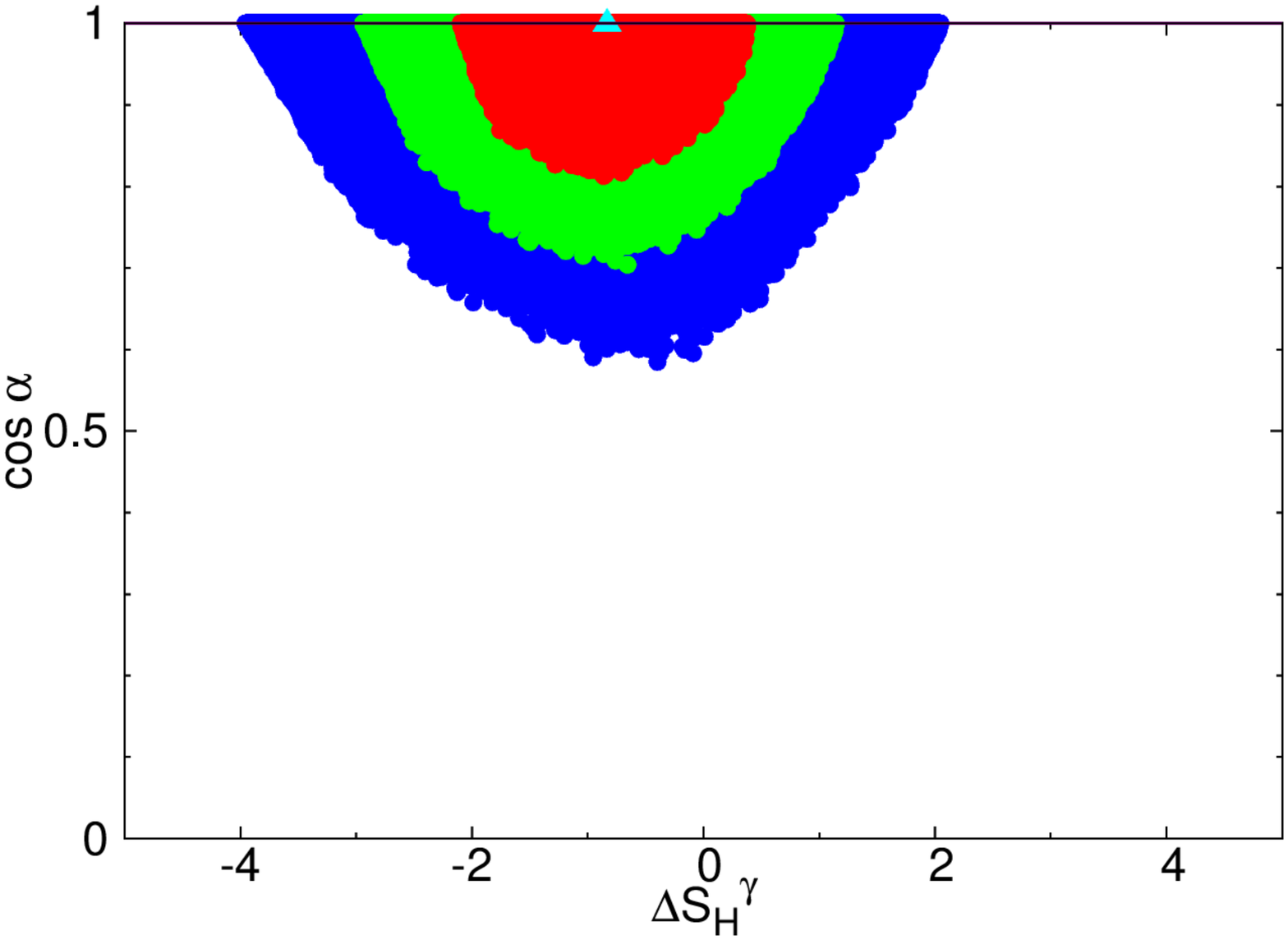}
\includegraphics[height=2.0in,angle=-90]{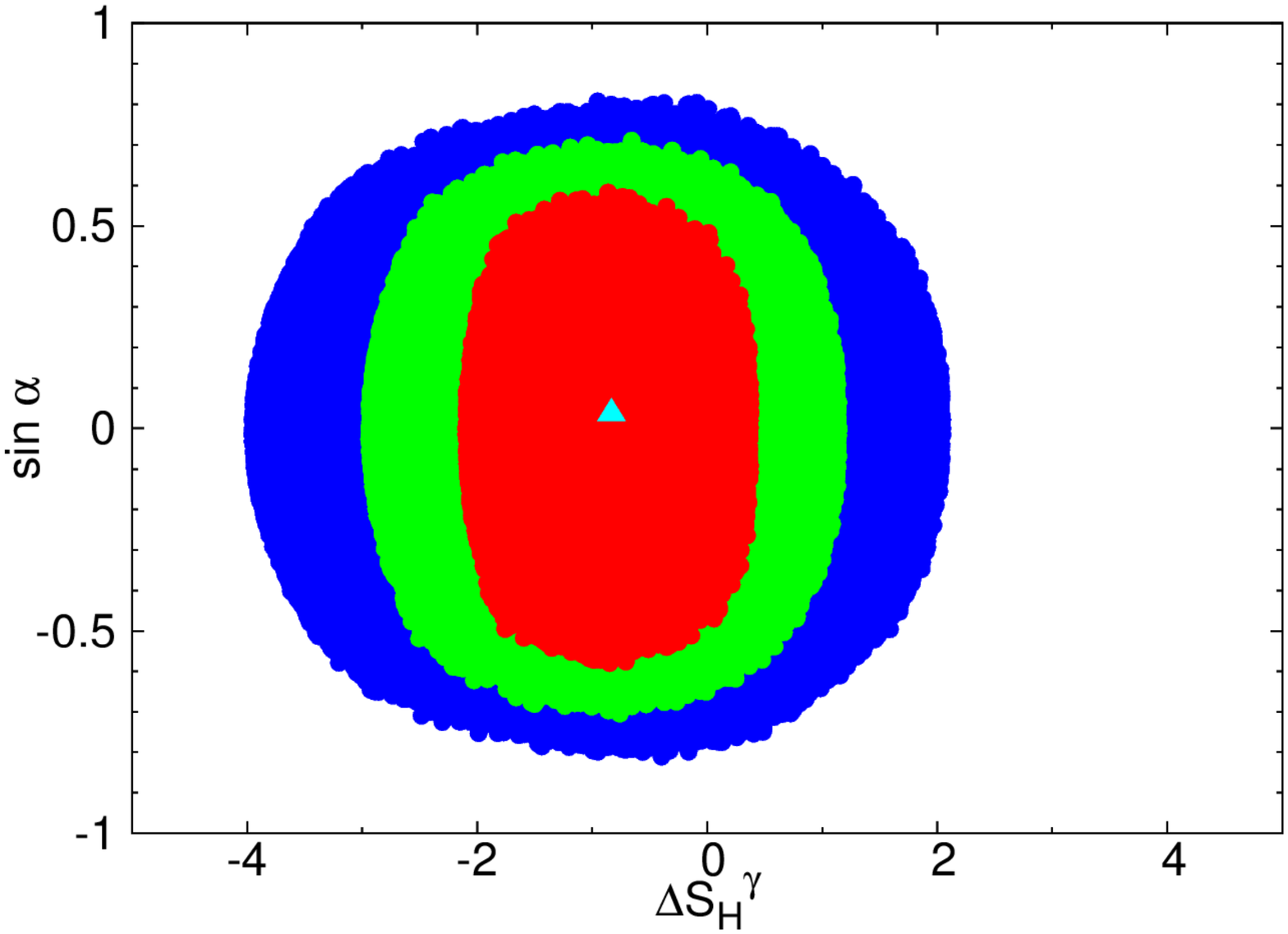}
\includegraphics[height=2.0in,angle=-90]{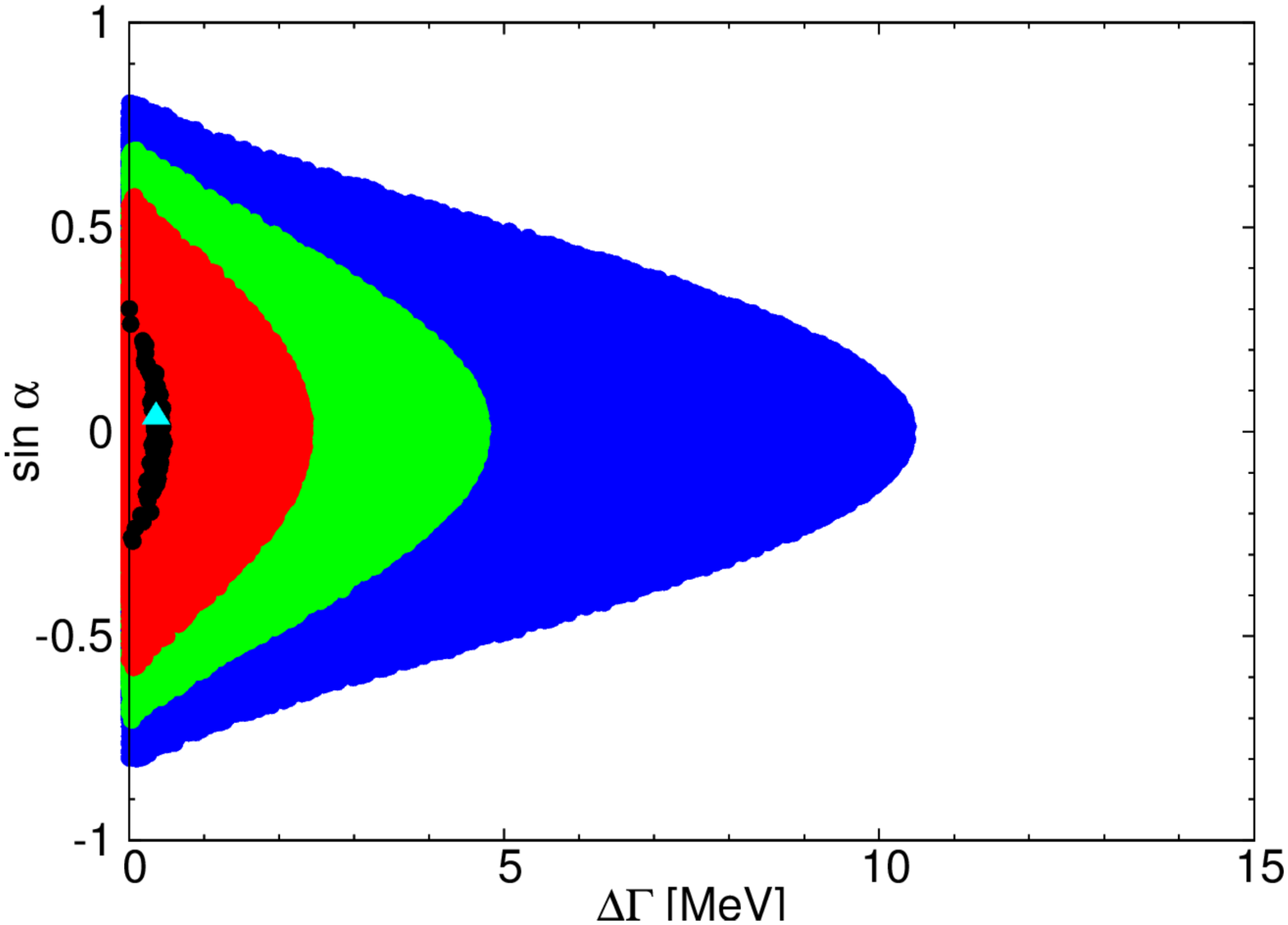}
\includegraphics[height=2.0in,angle=-90]{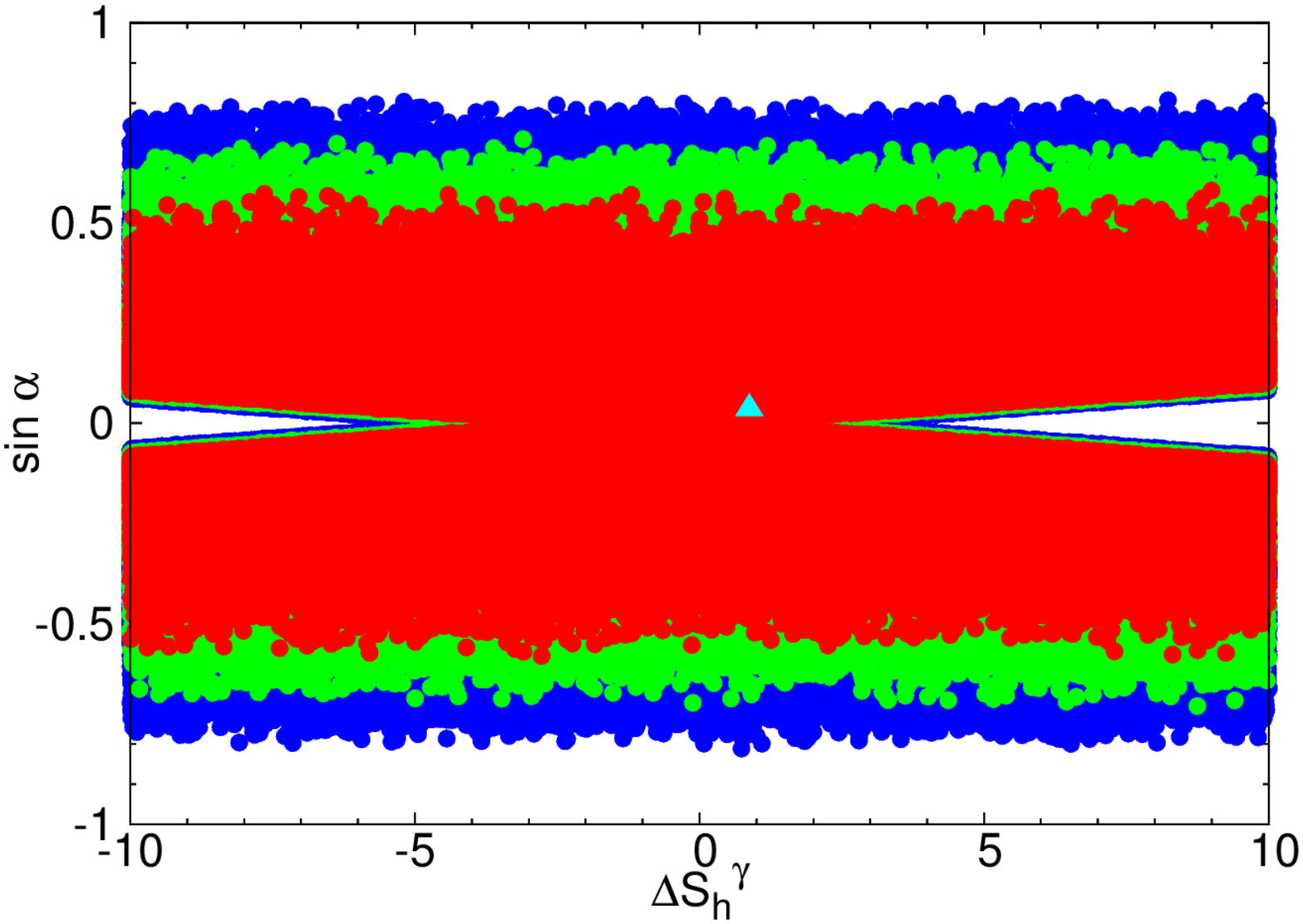}
\includegraphics[height=2.0in,angle=-90]{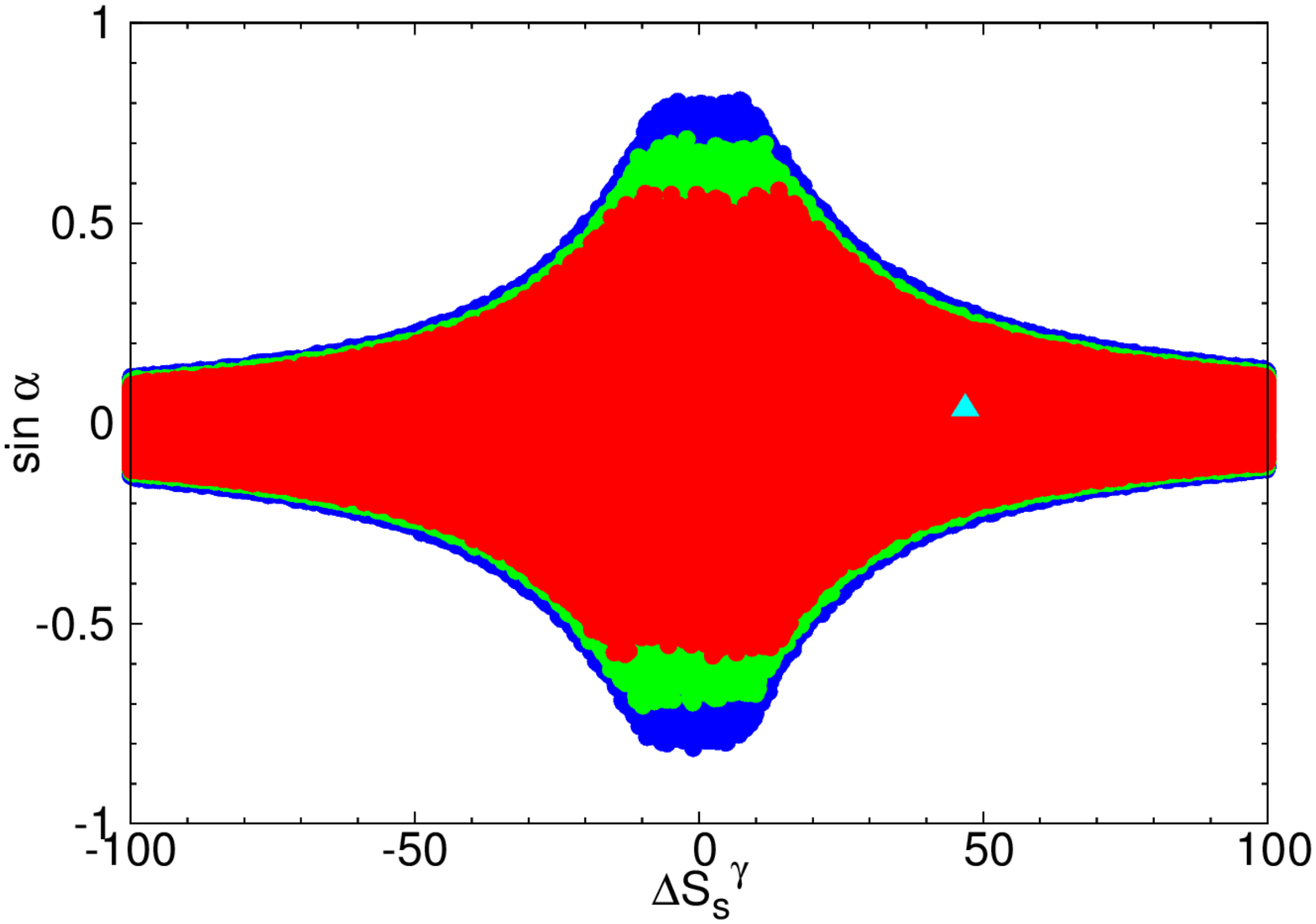}
\includegraphics[height=2.0in,angle=-90]{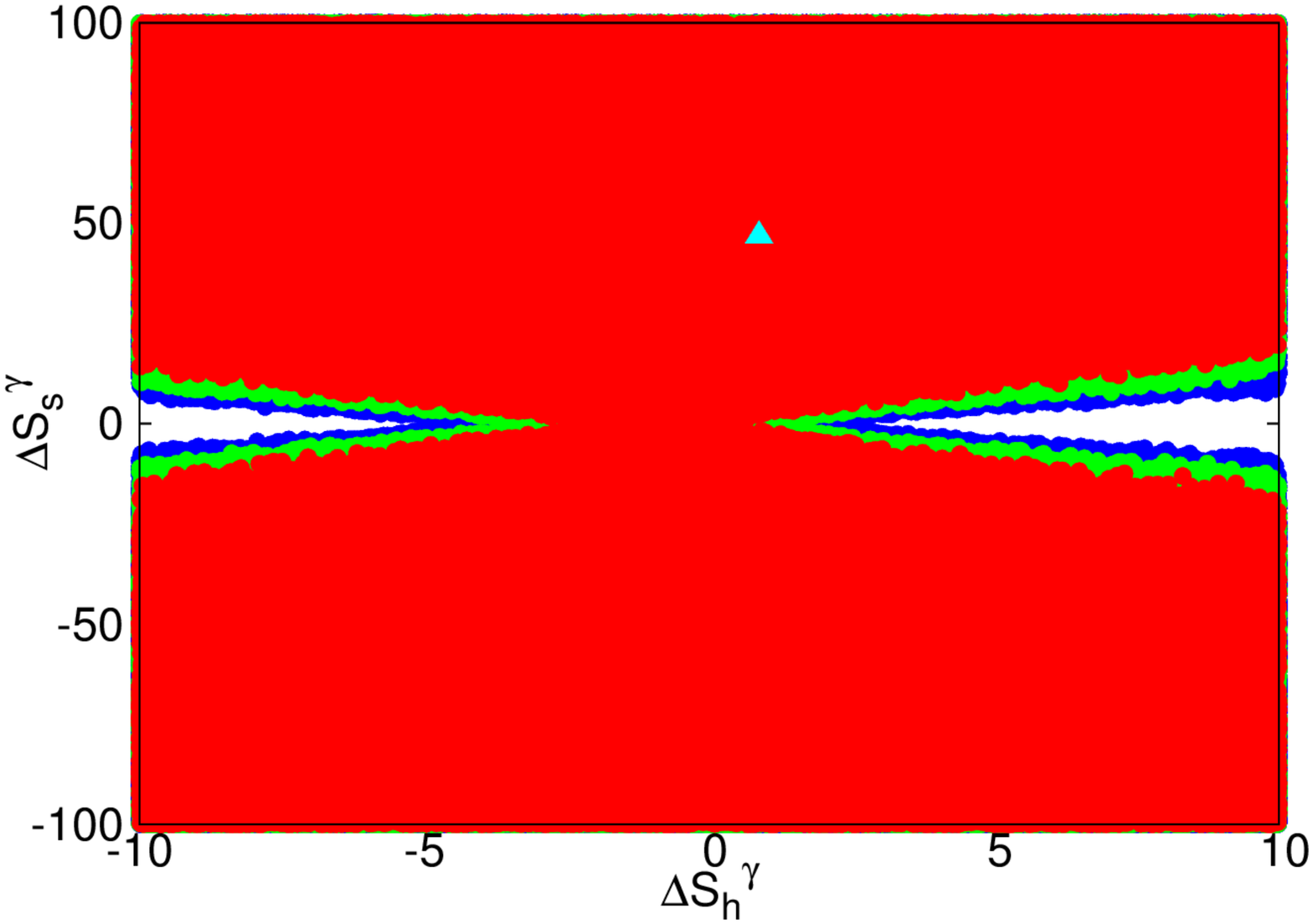}
\includegraphics[height=2.0in,angle=-90]{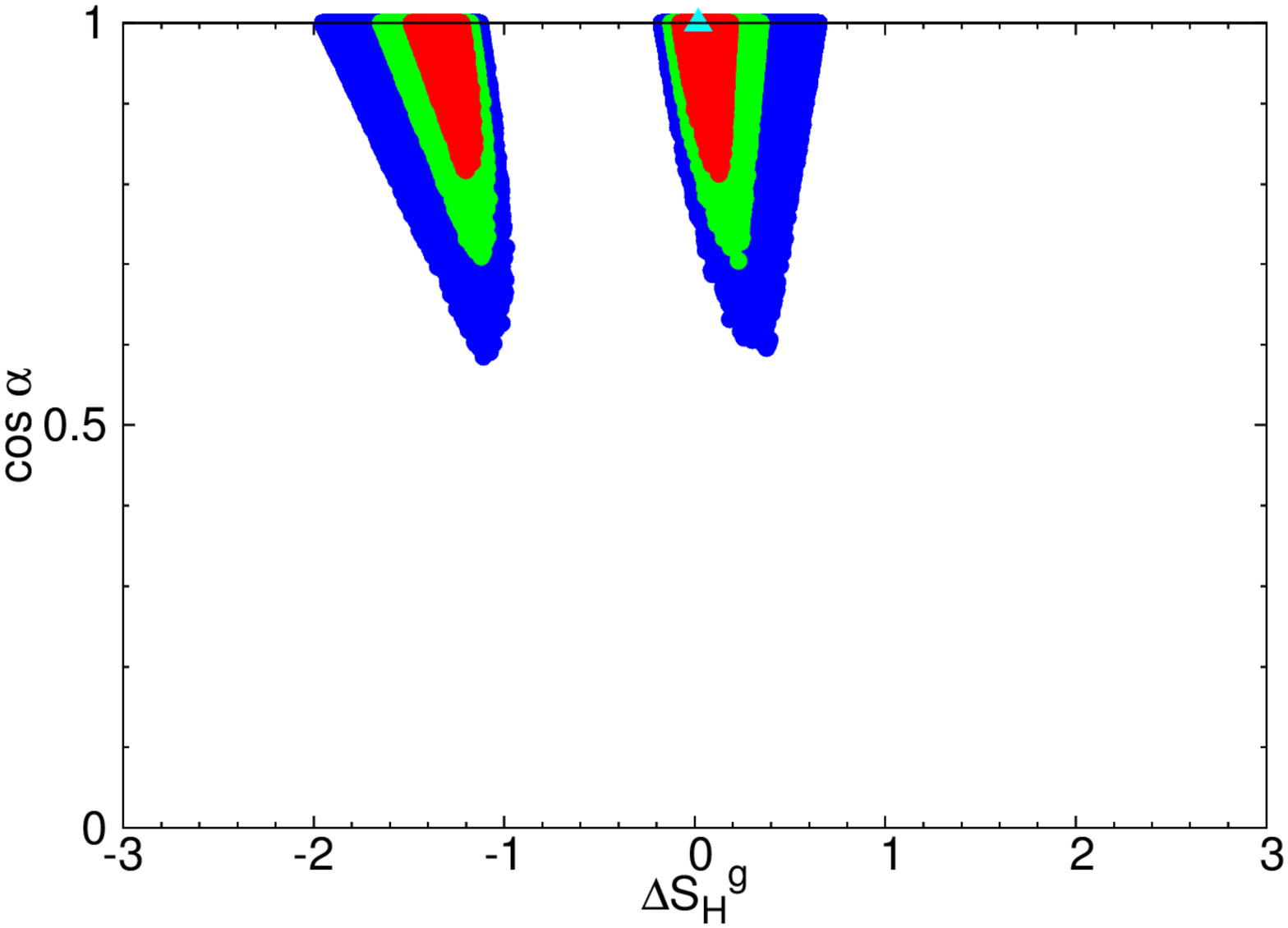}
\includegraphics[height=2.0in,angle=-90]{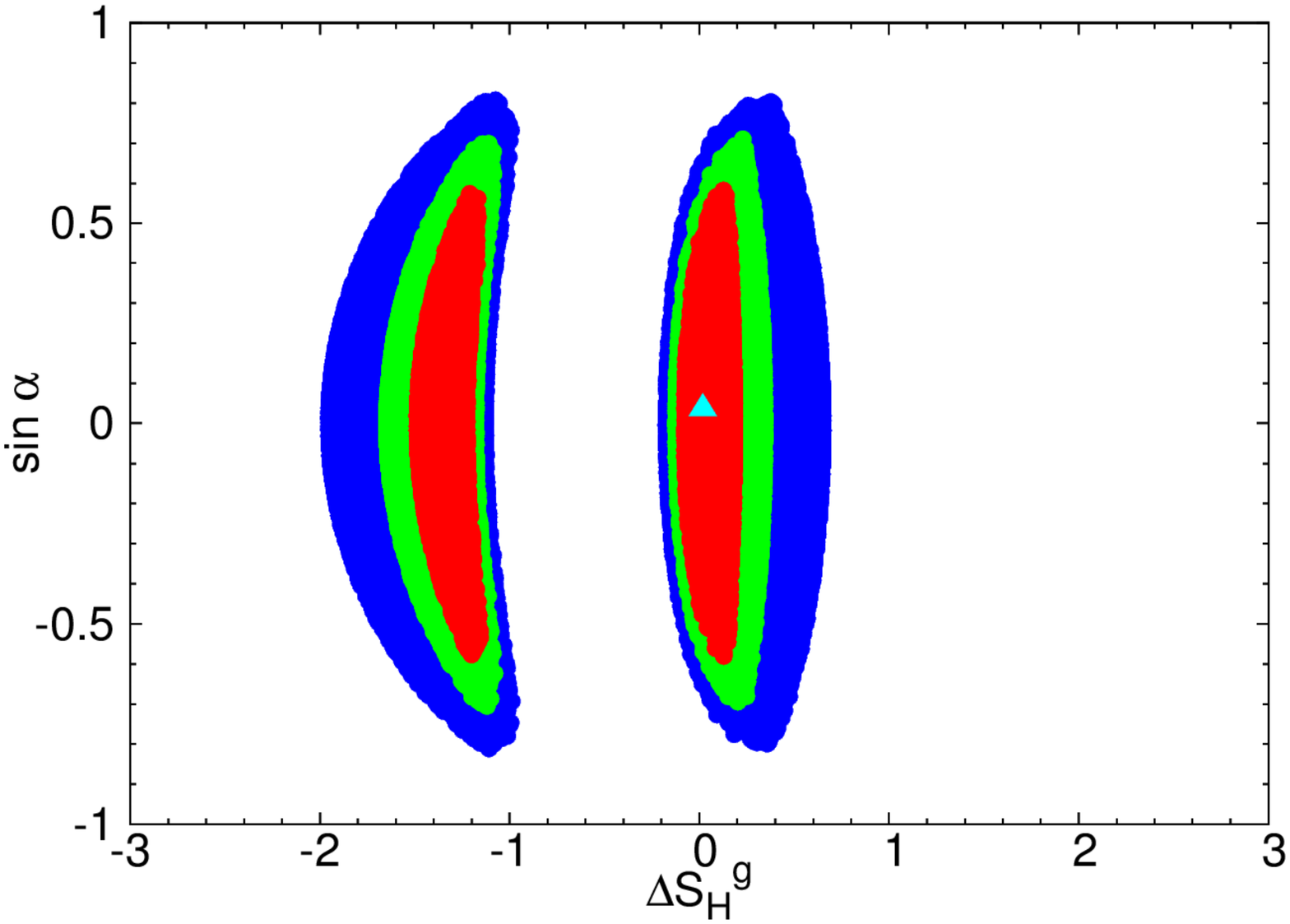}  \\
\includegraphics[height=2.0in,angle=-90]{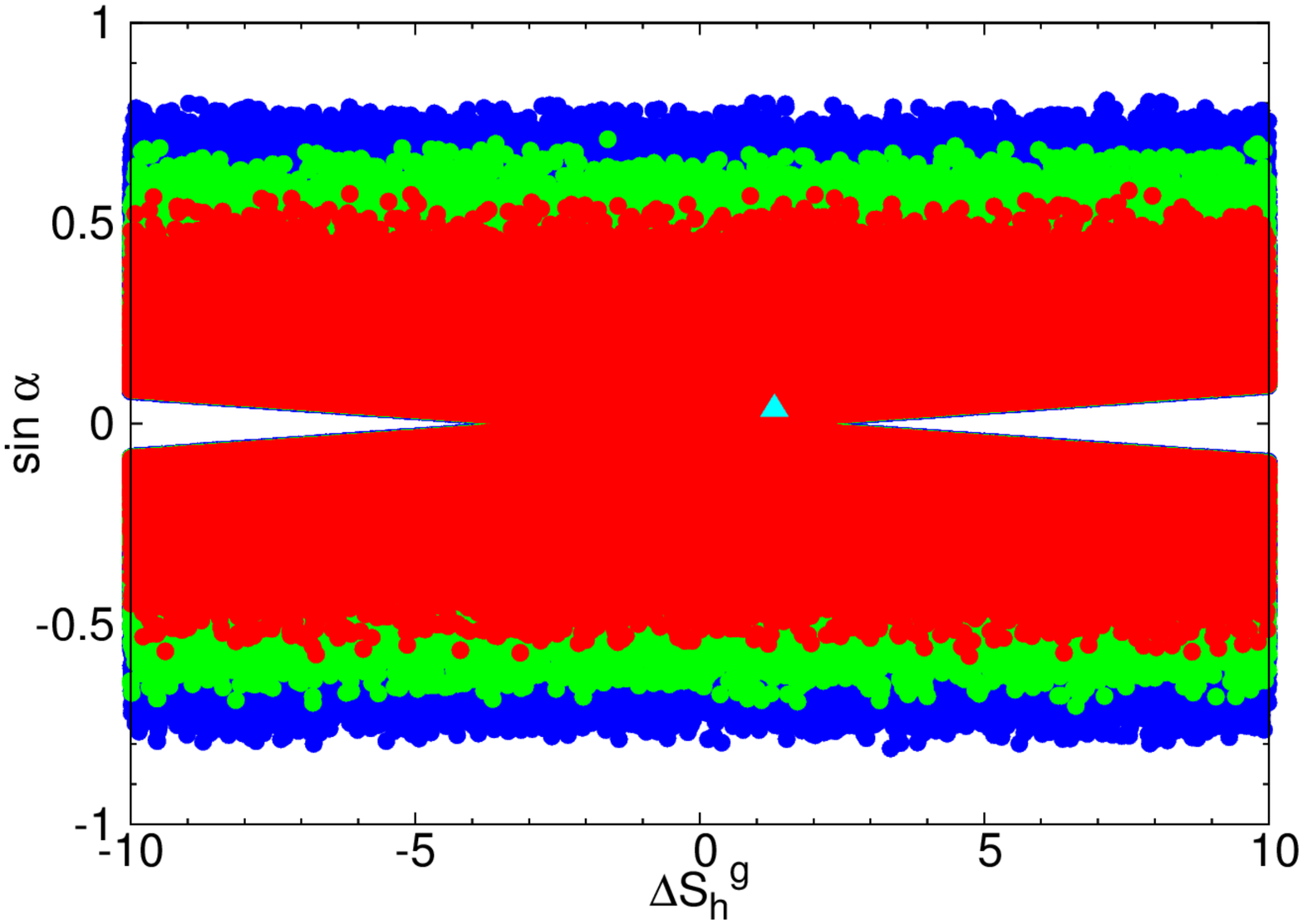}
\includegraphics[height=2.0in,angle=-90]{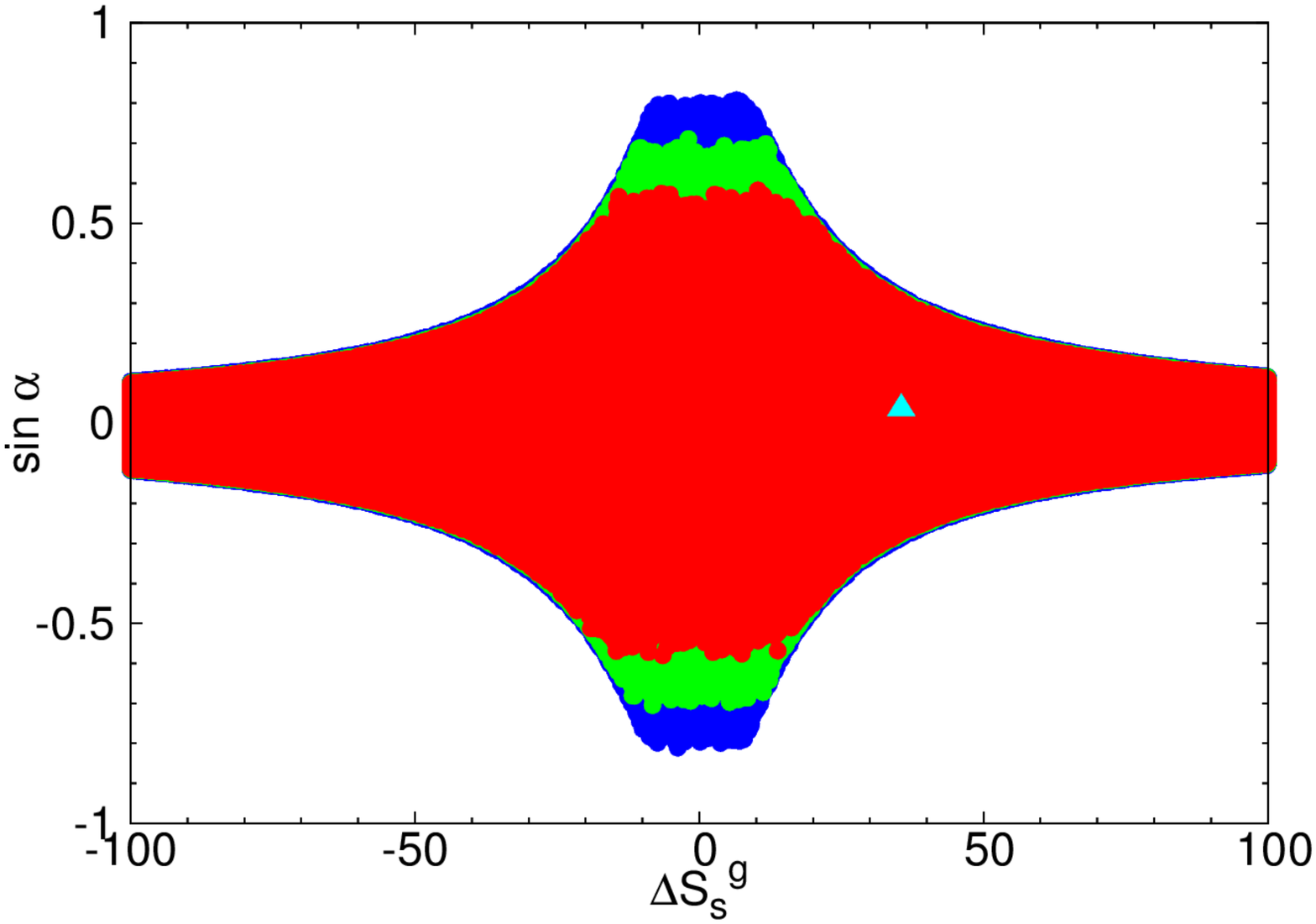}
\includegraphics[height=2.0in,angle=-90]{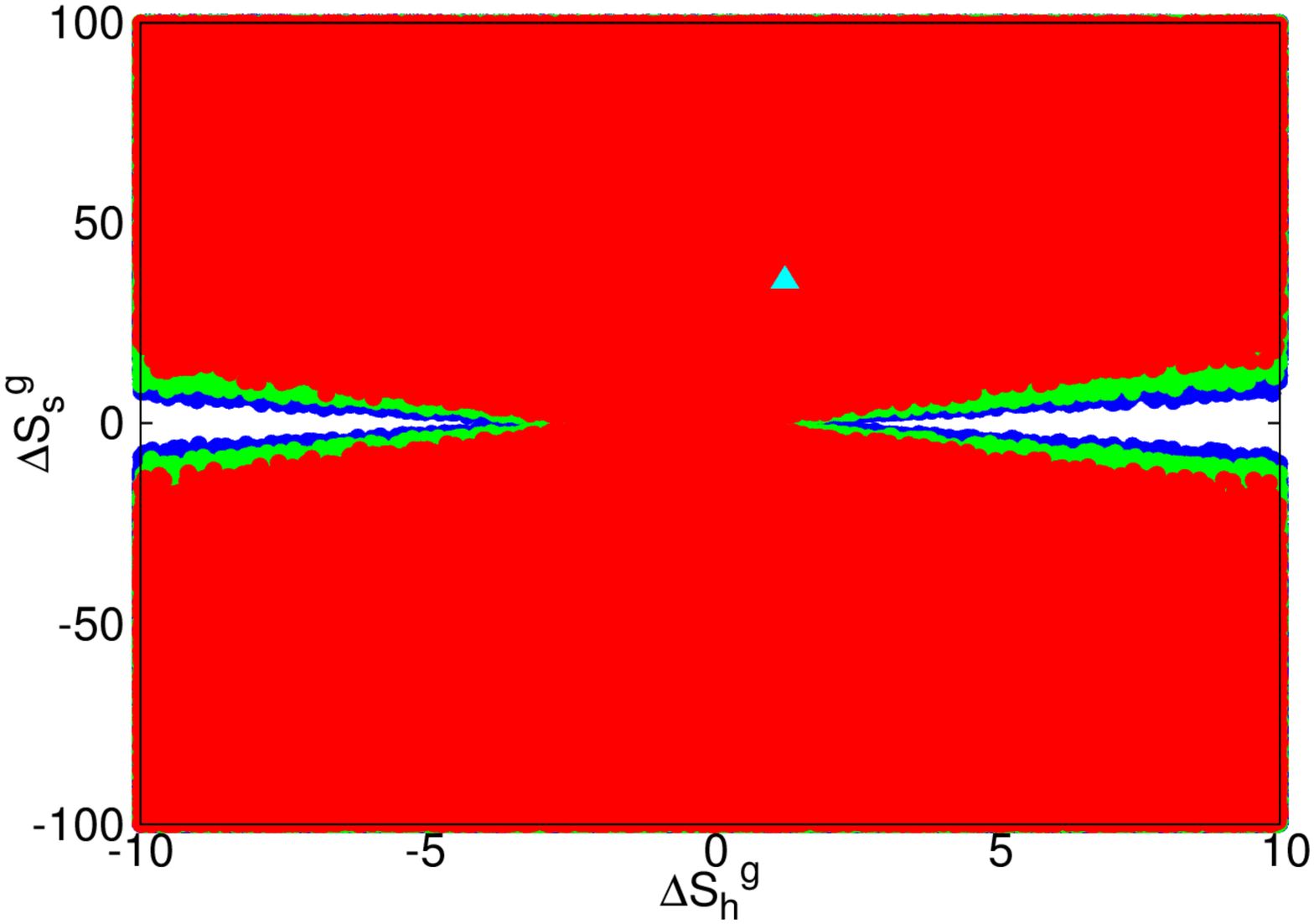}
\caption{\small \label{fig:sq}
The CL regions  for the {\bf SQ} fit.
The description of the CL regions is the same as in Fig.~\ref{fig:sd}.
}
\end{figure}

The CL regions are shown in Fig.~\ref{fig:sq}.
We observe that $\cos\alpha$ , $\sin\alpha$, $\Delta\Gamma_{\rm tot}$,
$\Delta S^\gamma_H$,  and $\Delta S^g_H$,  are well bounded as:
\begin{eqnarray}
&& \cos\alpha \gsim 0.70\,(0.58) \ {\rm at} \ 
95\%\, (99.7\%) \ {\rm CL}\,; \nonumber  \\[2mm]
&& |\sin\alpha| \lsim 0.71\,(0.81)\, \ {\rm at} \ 
95\%\, (99.7\%) \ {\rm CL}\,; \nonumber  \\[2mm]
&& \Delta\Gamma_{\rm tot} \lsim 4.70\,(10.40)\,{\rm MeV} \ {\rm at} \ 
95\%\, (99.7\%) \ {\rm CL}\,; \nonumber \\[2mm]
&& -2.94\,(-3.96) \lsim \Delta S^\gamma_H \lsim 1.14\,(2.05)\,, \ {\rm at} \ 
95\%\, (99.7\%) \ {\rm CL}\,; \nonumber  \\[2mm]
&& -0.13\,(-0.18) \lsim \Delta S^g_H \lsim 0.35\,(0.65)\, \ {\rm and} \ 
\nonumber   \\[2mm]
&& -1.65\,(-1.96) \lsim \Delta S^g_H \lsim -1.08\,(-1.00)\, \ {\rm at} \ 
95\%\, (99.7\%) \ {\rm CL}\,. \nonumber  
\end{eqnarray}

One may make similar observations for $\Delta S^{\gamma,g}_h$ and 
$\Delta S^{\gamma,g}_s$
as in the {\bf SL} case.
The parameters $\Delta S^{\gamma,g}_h$ and 
$\Delta S^{\gamma,g}_s$ are, in general, not bounded.
When $\sin\alpha=0$, $\Delta S^{\gamma,g}_h=\Delta S^{\gamma,g}_H$  and so they
are bounded as $\Delta S^{\gamma,g}_H$.
When $|\sin\alpha|$ takes its largest values, $|\Delta S^{\gamma,g}_s| \lsim 10$.
We also observe the wedges 
along $\sin\alpha=0$ and $\Delta S^{\gamma,g}_s=0$ due to the cancellation between
$\Delta S^{\gamma,g}_h\, \cos\alpha$ and $\Delta S^{\gamma,g}_s\, \sin\alpha$
when $|\Delta S^{\gamma,g}_h|>|\Delta S^{\gamma,g}_H|$.

The best-fit values for $\Delta S^\gamma_h (\Delta S^g_h)$ and 
$\Delta S^\gamma_s (\Delta S^g_s)$
are $0.875 (1.315)$ and $46.84 (35.54)$, respectively, 
even though $\Delta\chi^2$ does not change much
in most regions of the parameter space. Considering
$S^\gamma_{\rm SM}=-6.64\,(S^g_{\rm SM}=0.65)$ 
and the best-fit value $\Delta S^\gamma_H=-0.832 (\Delta S^g_H=0.019)$,
let alone a certain level of cancellation,
it would be very difficult to achieve such large values for 
$\Delta S^{\gamma,g}_h$ and $\Delta S^{\gamma,g}_s$, unless
the vector-like  quarks are light, come in with a large multiplicity,
and/or their Yukawa couplings to the singlet scalar $s$ are strong.

Before closing this section, it is worth mentioning the experimental
constraints on vector-like quarks and leptons.
The vector-like quarks and leptons have been searched at the Tevatron
and at the LHC.  The current best limits on vector-like quarks
are from the ATLAS collaboration
with $20.3$ fb$^{-1}$ luminosity at $8$ TeV~\cite{Aad:2015kqa}. 
The limits range between $715$ GeV and $950$ GeV for up-type vector-like
quarks and those for down-type ones between $575$ GeV and $813$ GeV.
Considering these limits, we observe that one should have ${\cal O}(100)$
vector-like  quarks to accommodate the large best-fit values
of $\Delta S^{\gamma ,g}_s$ shown in Table~\ref{tab:fits},
assuming ${\cal O}(1)$ Yukawa couplings of vector-like quarks
to the singlet scalar $s$.

\section{Discussion}

The Higgs-portal model involving a mixing between the SM Higgs field
and an $SU(2)$ singlet scalar boson is indeed the simplest extension to
the SM Higgs sector, and gives rise to interesting phenomenology. 
In particular, this type of models can provide dark matter candidates,
which exist in the hidden sector and interact with the SM sector through
the mixing. Since it involves the mixing, so it will have non-negligible
effects on the SM Higgs boson properties. In this work, we have used
the most updated Higgs boson data from LHC@7 and 8TeV to obtain very useful
constraints on the models. In the simplest of this class of models  
 -- the singlet
with a dark matter candidate ({\bf SD}), the deviations from the SM Higgs 
couplings can be parameterized by the mixing $\cos\alpha$ and the 
deviation in the total decay width $\Delta \Gamma_{\rm tot}$. We found that
the {\bf SD} 
model does not provide a better fit than the SM, and thus we obtain
the 95\% CL on the parameters:
\begin{eqnarray}
 \cos\alpha  \gsim  0.86\,, \ \ \
 \Delta \Gamma_{\rm tot}  \lsim  1.24 \;{\rm MeV}\,.
\end{eqnarray}
When more exotic particles are involved in the hidden sector, for example
the vector-like leptons ({\bf SL}) or vector-like quarks ({\bf SQ}) in this
work, the $H\gamma\gamma$ and $Hgg$ vertices are modified non-trivially, and 
thus more parameters are involved. The constraints on $\cos\alpha$ and
$\Delta \Gamma_{\rm tot}$ become somewhat less restrictive than the {\bf SD} case
(at 95\%CL):
\begin{eqnarray}
{\bf SL}: & &\;\; \cos\alpha \gsim 0.83 ,\;\;\; \Delta \Gamma_{\rm tot} 
  \lsim 1.9\;{\rm MeV} \nonumber \\
{\bf SQ}: & & \;\; \cos\alpha \gsim 0.70 ,\;\;\; \Delta \Gamma_{\rm tot} 
  \lsim 4.7\;{\rm MeV} \nonumber 
\end{eqnarray}
The allowed ranges for other parameters can be found in the previous 
section.

We also offer the following comments on our findings:
%
\begin{itemize}


\item The SM gives the best fit in terms of $\chi^2/d.o.f.$ although 
the difference from other best fits ({\bf SD, SL, SQ}) 
are not statistically significant yet.

\item {\bf SD} : In this case, the best $\chi^2$ occurs in the 
unphysical region: either $c_\alpha >1$ or $\Delta \Gamma < 0$. 
If this is still the case in the future data, a large class of DM
models (Higgs-portal fermion or vector DM, and DM models with local 
dark gauge symmetries) and non-SUSY $U(1)_{B-L}$ models 
will be strongly disfavored.  However, the usual Higgs-portal scalar DM model 
with $Z_2$ symmetry without the extra singlet scalar may still be 
viable, since the Higgs signal strength in that model will be the same 
as the SM case.

\item {\bf SL} : 
This case corresponds to the vector-like leptons in the loop for
$H\rightarrow \gamma\gamma$.
We get a reasonably good fit. Nevertheless, we need a 
rather large value for $\Delta S^\gamma_s = -16.27$,
which might be possible only if the vector-like leptons are light,
they come in with a large multiplicity,
or the Yukawa couplings of the vector-like lepton to the singlet 
scalar $s$ is strong.

\item {\bf SQ} : This case corresponds to the vector-like quarks in the 
loop for $H\rightarrow \gamma\gamma$
and $H\rightarrow gg$.  We get a reasonably good fit.  
However, we need a rather large value for 
$\Delta S^\gamma_s = 46.84$, 
which might be possible only if the vector-like 
quarks are light, they come in  with a large multiplicity, 
or their Yukawa coupling is very large.

\item {\bf SL} and {\bf SQ} : 
Though the best-fit values for 
$\Delta S^{\gamma,g}_h$ and $\Delta S^{\gamma,g}_s$
are large,  those for 
$\Delta S^{\gamma,g}_H = 
\Delta S^{\gamma,g}_h \cos_\alpha -\Delta S^{\gamma,g}_s \sin_\alpha$
are only about $-0.8$ and $0.02$, respectively.

\end{itemize}

\section{Acknowledgments}
The work of P.K. was supported in part by the Korea Neutrino Research
Center (KNRC) which is established by the National Research Foundation of
Korea (NRF) grant funded by the Korea government (MSIP) (No. 2009-0083526)
and in part by the NRF Research Grant NRF-2015R1A2A1A05001869.
The work of J.S.L. was supported by
the National Research Foundation of Korea (NRF) grant
(No. 2013R1A2A2A01015406).
The work of K.C. was supported by the MoST of Taiwan under Grants 
No. NSC 102-2112-M-007-015-MY3.


\newpage

\end{document}